\documentclass[
superscriptaddress,
preprint,
 amsmath,amssymb,
 aip,
]{revtex4-2}

\usepackage{graphicx}
\usepackage{dcolumn}
\usepackage{bm,siunitx}
\usepackage{comment}

\usepackage{color}

\begin{document}

\title{
\textcolor{black}{Spin dependent bandgap renormalization and state filling effect in Bi$_2$Se$_3$ observed by ultrafast Kerr rotation
}
}

\author{Kazuhiro Kikuchi}
\email{Authors to whom correspondence should be addressed: s2320272@u.tsukuba.ac.jp}
\affiliation{Department of Applied Physics, Graduate School of Pure and Applied Sciences, University of Tsukuba, 1-1-1 Tennodai, Tsukuba 305-8573, Japan}
\author{Yu Mizukoshi}
\affiliation{Department of Applied Physics, Graduate School of Pure and Applied Sciences, University of Tsukuba, 1-1-1 Tennodai, Tsukuba 305-8573, Japan}
\author{Takumi Fukuda}
\affiliation{Department of Applied Physics, Graduate School of Pure and Applied Sciences, University of Tsukuba, 1-1-1 Tennodai, Tsukuba 305-8573, Japan}
\affiliation{Femtosecond Spectroscopy Unit, Okinawa Institute of Science and Technology Graduate University, 1919-1 Tancha, Onna, Okinawa, Japan}
\author{Paul Fons}
\affiliation{Department of Electronics and Electrical Engineering, School of Integrated Design Engineering, Keio University, 4-1-1 Hiyoshi, Kohoku district, Yokohama city 223-8521, Japan}
\author{Muneaki Hase}
\affiliation{Department of Applied Physics, Graduate School of Pure and Applied Sciences, University of Tsukuba, 1-1-1 Tennodai, Tsukuba 305-8573, Japan}

\date{\today}

\begin{abstract}
We investigate the ultrafast spin dynamics of the prototypical topological insulator Bi\textsubscript{2}Se\textsubscript{3} using time-resolved Kerr-rotation (polarization-change) measurements across near-infrared wavelengths.
The Kerr-rotation angle $\Delta \theta_{K}$ of Bi\textsubscript{2}Se\textsubscript{3} was found to significantly depend on photon energy around a resonance transition ($\sim 1.0\ \mathrm{eV}$) of bulk states, as well as the ellipticity of the pump light, in the presence of spin excitation. 
The observed photon-energy dependence of $\Delta \theta_{K}$ can be well simulated by assuming spin-dependent refractive-index changes in the presence of band-gap renormalization and state-filling effect upon photoexcitation.
Our study delivers comprehensive insights into the opto-spintronic properties of bulk Bi\textsubscript{2}Se\textsubscript{3} and the fundamental physical processes underlying polarization changes. 
These findings are expected to be crucial in developing ultrafast magneto-optical memory devices, which can perform read-and-write operations in the Terahertz regime.


\end{abstract}

\maketitle


\textcolor{black}{Topological insulators (TIs) have a semiconductor-like bulk band and spin-polarized gapless surface bands \cite{hasan2010colloquium}, resulting in exotic properties such as spin momentum locking \cite{xie2014orbital,cao2013mapping,hsieh2009tunable,Park2012chiralorbitalPRL} and back-scattering suppression by nonmagnetic impurities\cite{Sunghun2014robustprotectionPRL,roushan2009topological}.}
\textcolor{black}{These properties are expected to catalyze the development of opto-spintronics by controlling spin-polarized current using light ellipticity \cite{mciver2012controlNNanotech,kastl2015ultrafast,braun2016ultrafast}.}
\textcolor{black}{A circularly polarized laser pulse has angular momentum, giving rise to spin excitation during ultrafast time scales from sub-picosecond to nanosecond \cite{manipulation_revmodphys,coherentspin_review}.
}
\textcolor{black}{
Spin dynamics have been actively studied through time-resolved and angle-resolved photoemission spectroscopy \cite{kuroda2017ultrafastPRB,kuroda2016generationPRL,soifer2019bandresolvedPRL,ketterl2018origin,reimann2014spectroscopy,RHuber2018subcycleNature} and ultrafast Kerr rotation measurement \cite{DHsiehPRLselective, MCwangPRLunraveling,boschini2015SciRep,mondal2018topological} in various materials, such as weak ferromagnets\cite{Kimel2005Nature}, semiconductors\cite{JAPInAsKini2008,NatComPerovskite2020,kimel2001roomtemp,PRBWSeExciton2014}, and TIs\cite{DHsiehPRLselective,mondal2018topological,MCwangPRLunraveling}.
}
\textcolor{black}{
In particular, previous optical studies on TIs have focused on the relationship between spin excitation and topological surface states (TSS).
}
\textcolor{black}{
However, the influence of the bulk state on the observed spin signal is not well understood. This is due to the coexistence of surface and bulk states on topological insulator surfaces, which makes it difficult to unambiguously assess which spin contribution is responsible. This difficulty is particularly pronounced in single wavelength measurements, as it is challenging to relate the obtained signal to the band structure.
}



\textcolor{black}{Furthermore, upon photoexcitation, various kinds of light-induced phenomena in addition to spin excitation are involved, such as Coulomb interactions between electrons and holes, nonlinear optical effects, and lattice vibrations \cite{chernikov2015population,Fukuda2023HighDensity, iwasakiFukuda2023APL, fukuda2024coherent}.
}
\textcolor{black}{
Therefore, it is essential to discuss the spin contribution, including TSS and the bulk state, and disentangle it from various components induced by photoexcitation.
}
\textcolor{black}{
A possible strategy for extracting the spin excitation is Kerr rotation measurement at various wavelengths, which can relate spin dynamics to band structure. This method allows us to discuss the spin-dependent modulation of the band structure induced by many-body effects.
}

In this study, we examine time-resolved Kerr rotation spin measurements at various wavelengths on a typical TI Bi$_2$Se$_3$ single crystal [Figure 1(a)] in order to clarify the relationship between spin-excitation and the bulk band structure. It is found that the photon-energy dependence of the Kerr rotation signal can be described by a spin-dependent energy shift and \textcolor{black}{decrease of the transition probability (state-filling effect; SFE)}\cite{NatComPerovskite2020}, which is governed by transitions between bulk states. As for the origin of the energy shift, we discuss the possibility of spin-dependent bandgap-renormalization (BGR) in the bulk state,\cite{PRBGiantSSBGR,PRBspindepGaAs,JAP2014spindepCdTe} by which the electronic properties of TIs can be modulated in a spin-selective way on ultrafast time scales. Thus, this study also offers a way for ultrafast modulation of the spin-polarized current of TIs via many-body effects. Moreover, we offer a paradigm for the precise measurement of spin dynamics in TSS by subtracting the bulk spin contribution from the Kerr rotation signal.

The sample \textcolor{black}{used in this study} was a bulk single crystal of (111) surface Bi$_2$Se$_3$ (HQ graphene) \textcolor{black}{with the thickness of $\sim 100\ \mathrm{\mu m}$}. \textcolor{black}{The sample surface was cleaved by an adhesive tape before experiments.}
\textcolor{black}{The orientation of the crystal axes was evaluated by coherent phonon measurements (see supplementary information).
}\textcolor{black}{
\textcolor{black}{To investigate the optical properties}, spectroscopic ellipsometry measurements were carried on for the sample by UVISEL PLUS (HORIBA). The measured real ($n$) and imaginary ($\kappa$) parts of the complex refractive index ($\hat{n}$) of $\mathrm{Bi_{2}Se_{3}}$ are shown in Fig. 1(b). The experimental data is well fit by the Lorentz model:
\begin{equation}
    \hat{\varepsilon}_{\mathrm{eq}}(E) = \varepsilon_{b}+ \sum_{j} \frac{C_j}{E^2-{E_{0j}}^2 - i E\gamma_j}.
\end{equation}
Here, $\hat{\varepsilon}_{\mathrm{eq}}(E)$ is a complex dielectric function for equilibrium conditions, $E$ is the photon energy in eV, $C_j$ is the amplitude of  $j$-th resonance, $E_{0j}$ is $j$-th resonant energy, $\gamma_j$ is  $j$-th linewidth, and $\varepsilon_b$ is the high-frequency limit of the dielectric function. The parameters of the Lorentz model are shown in Table 1, which are roughly consistent with a previous study\cite{eddrief2016bulktoulthin}. The bulk states were excited resonantly by 1.0 eV light because the resonant peak originates from bulk states\cite{eddrief2016bulktoulthin}.} 
\textcolor{black}{
Figure 1(c) shows possible optical transitions at 1.0 eV around the zone center and the zone boundary. 
The band structure of bulk $\mathrm{Bi_2Se_3}$ was calculated using the plane-wave density-functional theory program VASP 6.4.2. \cite{Kresse1993DFT}. Projector augmented waves forms of the pseudopotentials were used \cite{Kresse1999PAW}. The generalized gradient approximation (GGA) in the form of PBESol (PBE) was employed \cite{Perdew2008PBE}. A value of 520 eV was used for the energy cutoff based on a convergence study that yielded a tolerance of 1 meV in total energy. An 10×10×10 Monkhorst-Pack grid was used for Brillouin zone integration. Spin-orbit coupling was enabled. }

Around these points, transitions by circularly polarized light are allowed in $\mathrm{Bi_2Se_3}$ and similar topological insulator $\mathrm{Bi_{2}Te_{3}}$ because most of the upper edge of the valence band consists of electronic states whose total angular momentum is  $J=\pm3/2$ and most of the lower edge of conduction band consists of $J=\pm1/2$ states\cite{boschini2015SciRep,strongFaradayScirep}. \textcolor{black}{Using right circularly polarized light, electrons are excited from $J=-3/2$ states to $J=-1/2$ states while electrons are excited from $J=+3/2$ to $J=+1/2$ states by left circularly polarized light.} This trend is also observed in some semiconductors with strong spin-orbit coupling\cite{kimel2001roomtemp}.

\begin{table}[]
\begin{ruledtabular}
\begin{tabular}{ccccc}
        $j$ & 1 & 2 & 3 & 4 \\ \hline
        $C_j$ [eV$^2$]&     1.68 & 1.79 & 8.05 & 62.8\\
        $E_{0j}$ [eV]&0.67 & 1.01 & 1.44 & 2.04\\
        $\gamma_j$ [eV]&0.39 & 0.18 & 0.63 & 0.92\\
        $\varepsilon_b$ & \multicolumn{4}{c}{11.6}\\
\end{tabular}
    \caption{The fitted coefficients of the Lorentz model in Figure 1(b). }
\end{ruledtabular}
\end{table}

Time-resolved pump-probe Kerr rotation (TRKR) \textcolor{black}{measurements were carried out in reflection geometry under ambient conditions [Figure 2(a)]}.
A wavelength-variable near-infrared (NIR) pulse laser was used as the light source to explore the relation between spin-excitation and the bulk state. 
Femtosecond laser pulses from a Ti:Sapphire oscillator (Mantis, average power 500 mW, central wavelength 800 nm, pulse duration $\sim$ 20 fs, repetition rate 80 MHz) were amplified by 
a regenerative amplifier system (RegA9040) to obtain pulses with an average power of 500 mW, 
a central wavelength of 800 nm, a pulse duration of 40 fs, and a repetition rate of 100 kHz. 
These pulses were used for the optical parametric amplifier (OPA) to generate tunable light in the range of 1150-1500 nm (0.8 - 1.08 eV) with a pulse duration of $\sim \mathrm{60\ fs}$. 
The penetration depth for the use of the NIR wavelength range is estimated to be more than $\mathrm{100\ nm}$  a distance significantly larger than the $\sim 1$ nm depth in which TSS is thought to exist\cite{zhang2010crossover}. \textcolor{black}{As the experiment was carried out in air, the formation of a bismuth surface layer may occur. However, it is not expect to play a role regarding the spin physics in the bulk because the thickness of the bismuth layer is thin: The thickness of developed surface bismuth layer was measured to be $\sim 0.8$ nm\cite{edmonds2014stability}, a value much smaller than the penetration depth of the infrared light ($\sim 100$ nm).}

The generated NIR beam was split into an intense pump and weak probe beams. 
Both beams were focused onto the sample to a spot diameter of $\sim$ 70 $\mu \mathrm{m}$. Their incident angles were aligned normally to the sample surface to satisfy the assumptions of the following theoretical considerations\cite{FerromagmattHandbook,MagnetoOptics}. The polarization of the pump pulse was controlled by a quarter wave plate. \textcolor{black}{Reflected $p$-polarized probe light was split into $+\ang{45}$ and $-\ang{45}$ polarized light by a polarizing beam splitter, and detected by balanced InGaAs-PIN photodiodes (Hamamatsu Photonics, G12180-003A). The intensity difference in the split probe beam is proportional to the Kerr rotation of the probe pulse ($\Delta\theta_K$).} To obtain the time dependence of the Kerr rotation signal, the time delay between pump and probe pulses modulated at 9.5 Hz.



Figure 2\textcolor{black}{(b)} shows a typical TRKR measurement result induced by left and right circularly polarized pump pulses at 0.91 eV with a incident fluence of 0.5 mJ/cm$^2$.  
The sign of the transient zero-time delay peak reversed depending on the helicity of the pump pulse, which is typical for optical spin-excitation\cite{JAPInAsKini2008,PRBWSeExciton2014,DHsiehPRLselective,MCwangPRLunraveling,Kimel2005Nature}. The duration of the peak is nearly equal to the cross-correlation of the pump and probe pulses ($\approx$70 fs). Such spin excitation in topological insulators has been reported\cite{DHsiehPRLselective,MCwangPRLunraveling}. \textcolor{black}{In general, the Kerr rotation signal consists of spin-dependent and spin-independent terms\cite{mondal2018topological}. That is, the time trace of the Kerr rotation signal are \textcolor{black}{written as follows:}
\begin{equation}
    \Delta\theta_{K}^{\pm} = \pm f(t) + g(t).
\end{equation}
Here, $f(t)$ is the spin-dependent component, $g(t)$ is the spin-independent component (background), and $\pm$ refers to the helicity of the pump pulse. To eliminate the spin-independent background ($g(t)$) and verify the helicity-dependent component ($f(t)$), the difference between $\Delta\theta_{K}^{+}$ and $\Delta\theta_{K}^{-}$ was calculated [Fig. 2(b)]. 
As can be seen in Fig. 2(c), the spin-dependent term forms the large negative peak at zero time delay, with the area of the peak being the Kerr rotation intensity. }

To gain insight into the relationship between the spin excitation and the band structure, TRKR measurements were carried out at various wavelengths. 
\textcolor{black}{The photon-energy dependence of $\Delta\theta_{K}^{+}-\Delta\theta_{K}^{-}$ for an incident pump fluence of \si{mJ/cm^2} is shown in Fig. 3(a). According to the measured reflectivity spectrum (not shown), the variation of absorbed fluence over the range of photon energy used was less than 2\%, a value much smaller than the change of the Kerr rotation signal shown in Fig. 3. Therefore,  it can be assumed that the absorbed fluence remains nearly constant over the photon energy range used.}
In the following discussion, the Kerr rotation intensity data was derived from the data shown in Fig. 3(a). In Fig. 3(a), a minor oscillatory component at 4.2 THz was observed. This corresponds to Raman active $E_{g}^{2}$ coherent phonon mode\cite{BiSeRamanNanolett2011,norimatsu2013SSC}, which is occasionally observed in Kerr rotation and is excited by a circularly polarized pump\cite{norimatsu2015allphonon}. Because the amplitude of the oscillation is nearly at the noise limit, a clear trend of photon-energy dependence was not observed. 

\textcolor{black}{Conventionally, spin excitation by circularly polarized light has been phenomenologically described by the inverse Faraday effect, resulting in the changes of Kerr-rotation signals in the time domain\cite{Kimel2005Nature, manipulation_revmodphys}.} \textcolor{black}{However, to describe the photon-energy dependence of the Kerr rotation, a more detailed description of the Kerr rotation signal is required. Therefore}, we introduce a description of the Kerr rotation using the optical constants of the sample\cite{FerromagmattHandbook,MagnetoOptics}:
\begin{equation}\label{Kerrrotation_kappa}
    \Delta\theta_{K} = -\mathrm{Im}\left[\frac{\hat{n}_{+}-\hat{n}_{-}}{\hat{n}_{+}\hat{n}_{-}-1}\right].
\end{equation}
Here, $\hat{n}_{\pm}$ is the complex refractive index for right/left circularly polarized light, assuming that the difference in $\hat{n}_{\pm}$ is sufficiently small $(\hat{n}_{+}+\hat{n}_{-} \gg \hat{n}_{+}-\hat{n}_{-})$ and the linearly polarized probe light incidents normal to the surface of the sample. In our setup, $\hat{n}_{+}$ corresponds to the transition from $J=-3/2$ to $J=-1/2$ and $\hat{n}_{-}$ corresponds to the transition from $J=+3/2$ to $J=+1/2$. \textcolor{black}{As shown in Eq. (\ref{Kerrrotation_kappa}), the difference in $\hat{n}_{\pm}$ leads to the nonzero Kerr rotation. On the contrary, under equilibrium, $\hat{n}_{+}$ and $\hat{n}_{-}$ are equal and the Kerr rotation is zero. The possible reasons for the difference in $\hat{n}_{\pm}$ are spin-dependent quantum-mechanical phenomena such as spin-dependent level shifts and differences in the transition probability for right/left circularly polarized light. }

From here, we discuss the possible origins of the spin-dependent level shift and changes in the transition probability, 
\textcolor{black}{both of which modify the conventional dielectric function (Eq. 1) as explained below.}
First, we discuss the former, spin-dependent energy shift. The shift has been attributed to (I) the optical Stark effect by circularly polarized light\cite{OSENatMat2015,OSEprobedKerr_PRB} or (II) spin-dependent band gap renormalization (BGR)\cite{PRBGiantSSBGR,PRBspindepGaAs,JAP2014spindepCdTe}. However, in our experiment, photoexcited carriers should exist because the sample was pumped by the above bandgap light. Therefore, (I) the optical Stark effect can be safely ruled out, and (II) band-gap re-normalization is a more plausible origin. It is well known that band gaps in semiconductors are \textit{renormalized} by many-body effects arising from the presence of free carriers. In particular, screening of exchange-correlation due to the presence of photogenerated carriers in the system reduces the band gap, referred to as the BGR effect\cite{BGRNatPhoto2015}. In spin-dependent BGR, one of the degenerated energy levels is selectively populated by circularly polarized light\cite{PRBGiantSSBGR}, and thus, an energy shift occurs. 
\textcolor{black}{In order to take into account the BGR which causes a reduction in the resonant energy $E_{0j}$, $E_{0j}$ in Eq. (1) was changed to $(E_{0j}-\Delta E)$, where $\Delta E$ refers to the magnitude of the energy shift.}

Next, we discuss spin-dependent changes in the transition probability. This phenomenon is well known as the SFE \cite{BGRNatPhoto2015}. When the sample is pumped by intense light, the transition probability decreases due to the Pauli-blocking. Such phenomena can occur in a spin-dependent way\cite{PRBspindepGaAs}. To include the effects of SFE, the complex dielectric function($\hat{\varepsilon}_{\mathrm{eq}}( E)$) in \textcolor{black}{Eq. (1)} was modified. Due to the SFE, the transition probability decreases, which results in a decrease of $C_j$\cite{HorKerrPRB2013,kimel2001roomtemp}. This is because $C_j$ is proportional to the transition probability. 
\textcolor{black}{To account for the SFE, $C_j$ in Eq. (1) is replaced by $A\times C_j$, where $A$ is a constant $(0<A<1)$.}
In this expression, it was assumed that the oscillator strength decreases monotonically. This is because the wavelength of the pump and probe beam are equal in our experiment; thus, the SFE should be observed at any wavelength. From the above discussion, with the assumption that SFE and spin-dependent BGR occur simultaneously, the dielectric function was modified to be:
\begin{equation}
    \hat{\varepsilon}'( E) = 
    \varepsilon_{b}+ A \sum_{j} \frac{C_j}{E^2- (E_{0j}-\Delta E)^2 - i E\gamma_j}.
\end{equation}
\textcolor{black}{As mentioned above, $\Delta E$ and $A$ represent the magnitudes of BGR and SFE, respectively. By excitation of the spin-dependent BGR with right circularly polarized light, the energy difference between the $J=-3/2$ and $J=-1/2$ states decreases.}

Therefore, when the sample is excited by right/left circularly polarized light, the refractive index can be described by $\hat{n}_{\pm}( E)=\hat{\varepsilon}'( E)^2$ and $\hat{n}_{\mp}( E)=\hat{\varepsilon}_{\mathrm{eq}}( E)^2$. Considering that the Kerr rotation intensity was derived from the subtracted signal, the photon-energy dependence of Kerr rotation intensity $\Delta\theta_{K}( E)$ is described by
\begin{align}
    \Delta\theta_{K}( E)
        &\propto
            -\mathrm{Im}\left[\frac{\hat{n}_{+}-\hat{n}_{-}}{\hat{n}_{+}\hat{n}_{-}-1}\right]_{+}
            -\left(-\mathrm{Im}\left[\frac{\hat{n}_{+}-\hat{n}_{-}}{\hat{n}_{+}\hat{n}_{-}-1}\right]_{-}\right)  \\
        &=  -\mathrm{Im}\left[\frac{\hat{\varepsilon}'( E)^2-\hat{\varepsilon}_{\mathrm{eq}}( E)^2}{\hat{\varepsilon}'( E)^2\hat{\varepsilon}_{\mathrm{eq}}( E)^2-1}\right]
            +\mathrm{Im}\left[\frac{\hat{\varepsilon}_{\mathrm{eq}}( E)^2-\hat{\varepsilon}'( E)^2}{\hat{\varepsilon}_{\mathrm{eq}}( E)^2\hat{\varepsilon}'( E)^2-1}\right]  \\
        &= -2\ \mathrm{Im}\left[\frac{\hat{\varepsilon}'( E)^2-\hat{\varepsilon}_{\mathrm{eq}}( E)^2}{\hat{\varepsilon}'( E)^2\hat{\varepsilon}_{\mathrm{eq}}( E)^2-1}\right].
\end{align}
Here, the first term in \textcolor{black}{Eq.} (7) refers to the Kerr rotation intensity spectrum for the right circularly polarized pump, while the second term refers to that for the left circularly polarized light. $\hat{\varepsilon}'( E)$ is the modified dielectric function in \textcolor{black}{Eq.} (6). As shown in \textcolor{black}{Fig.} 3(b), a fit using \textcolor{black}{Eq.} (9) reasonably reproduces the experimental result. In this fitting procedure, the fitting coefficients in Table 1 were used for the parameters in the dielectric function in conjunction with the free parameters $A,\ \Delta E$, and the scaling factor. The parameter $A=0.95\pm0.01$ indicates that the transition probability decreases by 5 percent in a spin-selective way. Such several percent modulation is also observed in the form of modulation of the optical constant of $\mathrm{Bi_{2}Se_{3}}$ with  $\sim1\ \mathrm{mJ/cm^2}$ light\cite{shang2020saturable}. The parameter $\Delta E$ indicates an energy shift of about $76\pm10$ meV, which is the same order of the exciton binding energy of $\mathrm{Bi_2Te_3}$\cite{mori2023spatiallyNature} and  $\mathrm{Bi_{2}Se_{3}}$\cite{PRLScaling2017} but larger than these reports. To determine the magnitude of the energy shift more precisely, it is required to perform measurements over a wider wavelength range in the near future.

It is worth noting that the above discussion offers a pathway for precise measurement of spin dynamics in TSS. In the optical detection approach, researchers have struggled with extracting TSS contribution from measurement results which includes both contributions from the surface and bulk states. Moreover, because the bulk band structure exists even at the sample surface, surface-sensitive methods are not always TSS-sensitive. 
\textcolor{black}{To measure the spin dynamics in TSS precisely, it is necessary to carry out spin measurements on thin film samples (from $\sim5$ nm to several tens of nm), where the contribution from the bulk states decreases and the corresponding contribution from the TSS will be observed. Note that the dielectric function of the TSS was reported to be well described by the Drude model\cite{Ou2014ultravioletNatCom}. Therefore, if the TSS contribution is present, the photon energy dependent Kerr rotation signal will change to incorporate the Drude-like response and the effects of both the BGR and SFE will also changed.
}

\textcolor{black}{
In conclusion, we investigated the femtosecond spin dynamics of bulk Bi$_2$Se$_3$ using a time-resolved Kerr rotation technique for various photon energies from 0.8 to 1.08 eV. By subtracting the Kerr rotation signal for the left-circularly polarized pump from the right-circularly polarized pump, we obtained a photon-energy-dependent Kerr rotation signal, which is governed by the transitions between the bulk states at the zone center and zone boundary. 
The observed photon-energy dependence of the Kerr rotation signal can be well reproduced by incorporating both spin-dependent refractive-index changes in the presence of band-gap renormalization and the state-filling effect upon photoexcitation. 
Our study delivers a potential way for the precise measurement of spin dynamics in topological surface states by subtracting the bulk spin contribution from the Kerr rotation signal, which will be crucial in developing ultrafast magneto-optical memory devices, with capabilities extending to read-and-write operations in the Terahertz regime.
}

\newpage

\section*{Supplementary Material}
See the supplementary material for information about evaluation of the orientation of the crystal axes by coherent phonon measurements.

\section*{Acknowledgement}
This work was supported by JSPS Research Fellowships for Young Scientists, JSPS KAKENHI (Grant Numbers. 22H01151, 23K22422, and 22KJ0352), and CREST, JST (Grant Number. JPMJCR1875). We thank the Faculty of Pure and Applied Science and the Organization for Open Facility Initiatives, University of Tsukuba, for the measurement using a spectroscopic ellipsometer. TF acknowledges the support by the Sasakawa Scientific Research Grant from The Japan Science Society.
\section*{AUTHOR DECLARATIONS}
\section*{Conflict of Interest}
The authors have no conflicts to disclose.


\section*{DATA AVAILABILITY}
The data that support the findings of this study are available from the corresponding author upon reasonable request.

\bibliography{ms}

\providecommand{\noopsort}[1]{}\providecommand{\singleletter}[1]{#1}%
\begin{thebibliography}{54}%
\makeatletter
\providecommand \@ifxundefined [1]{%
 \@ifx{#1\undefined}
}%
\providecommand \@ifnum [1]{%
 \ifnum #1\expandafter \@firstoftwo
 \else \expandafter \@secondoftwo
 \fi
}%
\providecommand \@ifx [1]{%
 \ifx #1\expandafter \@firstoftwo
 \else \expandafter \@secondoftwo
 \fi
}%
\providecommand \natexlab [1]{#1}%
\providecommand \enquote  [1]{``#1''}%
\providecommand \bibnamefont  [1]{#1}%
\providecommand \bibfnamefont [1]{#1}%
\providecommand \citenamefont [1]{#1}%
\providecommand \href@noop [0]{\@secondoftwo}%
\providecommand \href [0]{\begingroup \@sanitize@url \@href}%
\providecommand \@href[1]{\@@startlink{#1}\@@href}%
\providecommand \@@href[1]{\endgroup#1\@@endlink}%
\providecommand \@sanitize@url [0]{\catcode `\\12\catcode `\$12\catcode `\&12\catcode `\#12\catcode `\^12\catcode `\_12\catcode `\%12\relax}%
\providecommand \@@startlink[1]{}%
\providecommand \@@endlink[0]{}%
\providecommand \url  [0]{\begingroup\@sanitize@url \@url }%
\providecommand \@url [1]{\endgroup\@href {#1}{\urlprefix }}%
\providecommand \urlprefix  [0]{URL }%
\providecommand \Eprint [0]{\href }%
\providecommand \doibase [0]{https://doi.org/}%
\providecommand \selectlanguage [0]{\@gobble}%
\providecommand \bibinfo  [0]{\@secondoftwo}%
\providecommand \bibfield  [0]{\@secondoftwo}%
\providecommand \translation [1]{[#1]}%
\providecommand \BibitemOpen [0]{}%
\providecommand \bibitemStop [0]{}%
\providecommand \bibitemNoStop [0]{.\EOS\space}%
\providecommand \EOS [0]{\spacefactor3000\relax}%
\providecommand \BibitemShut  [1]{\csname bibitem#1\endcsname}%
\let\auto@bib@innerbib\@empty
\bibitem [{\citenamefont {Hasan}\ and\ \citenamefont {Kane}(2010)}]{hasan2010colloquium}%
  \BibitemOpen
  \bibfield  {author} {\bibinfo {author} {\bibfnamefont {M.~Z.}\ \bibnamefont {Hasan}}\ and\ \bibinfo {author} {\bibfnamefont {C.~L.}\ \bibnamefont {Kane}},\ }\bibfield  {title} {\enquote {\bibinfo {title} {Colloquium: {T}opological {I}nsulators},}\ }\href@noop {} {\bibfield  {journal} {\bibinfo  {journal} {Rev. Mod. Phys.}\ }\textbf {\bibinfo {volume} {82}},\ \bibinfo {pages} {3045} (\bibinfo {year} {2010})}\BibitemShut {NoStop}%
\bibitem [{\citenamefont {Xie}\ \emph {et~al.}(2014)\citenamefont {Xie}, \citenamefont {He}, \citenamefont {Chen}, \citenamefont {Feng}, \citenamefont {Yi}, \citenamefont {Liang}, \citenamefont {Zhao}, \citenamefont {Mou}, \citenamefont {He}, \citenamefont {Peng} \emph {et~al.}}]{xie2014orbital}%
  \BibitemOpen
  \bibfield  {author} {\bibinfo {author} {\bibfnamefont {Z.}~\bibnamefont {Xie}}, \bibinfo {author} {\bibfnamefont {S.}~\bibnamefont {He}}, \bibinfo {author} {\bibfnamefont {C.}~\bibnamefont {Chen}}, \bibinfo {author} {\bibfnamefont {Y.}~\bibnamefont {Feng}}, \bibinfo {author} {\bibfnamefont {H.}~\bibnamefont {Yi}}, \bibinfo {author} {\bibfnamefont {A.}~\bibnamefont {Liang}}, \bibinfo {author} {\bibfnamefont {L.}~\bibnamefont {Zhao}}, \bibinfo {author} {\bibfnamefont {D.}~\bibnamefont {Mou}}, \bibinfo {author} {\bibfnamefont {J.}~\bibnamefont {He}}, \bibinfo {author} {\bibfnamefont {Y.}~\bibnamefont {Peng}}, \emph {et~al.},\ }\bibfield  {title} {\enquote {\bibinfo {title} {Orbital-selective spin texture and its manipulation in a topological insulator},}\ }\href@noop {} {\bibfield  {journal} {\bibinfo  {journal} {Nat. Commun.}\ }\textbf {\bibinfo {volume} {5}},\ \bibinfo {pages} {3382} (\bibinfo {year} {2014})}\BibitemShut {NoStop}%
\bibitem [{\citenamefont {Cao}\ \emph {et~al.}(2013)\citenamefont {Cao}, \citenamefont {Waugh}, \citenamefont {Zhang}, \citenamefont {Luo}, \citenamefont {Wang}, \citenamefont {Reber}, \citenamefont {Mo}, \citenamefont {Xu}, \citenamefont {Yang}, \citenamefont {Schneeloch} \emph {et~al.}}]{cao2013mapping}%
  \BibitemOpen
  \bibfield  {author} {\bibinfo {author} {\bibfnamefont {Y.}~\bibnamefont {Cao}}, \bibinfo {author} {\bibfnamefont {J.}~\bibnamefont {Waugh}}, \bibinfo {author} {\bibfnamefont {X.}~\bibnamefont {Zhang}}, \bibinfo {author} {\bibfnamefont {J.-W.}\ \bibnamefont {Luo}}, \bibinfo {author} {\bibfnamefont {Q.}~\bibnamefont {Wang}}, \bibinfo {author} {\bibfnamefont {T.}~\bibnamefont {Reber}}, \bibinfo {author} {\bibfnamefont {S.}~\bibnamefont {Mo}}, \bibinfo {author} {\bibfnamefont {Z.}~\bibnamefont {Xu}}, \bibinfo {author} {\bibfnamefont {A.}~\bibnamefont {Yang}}, \bibinfo {author} {\bibfnamefont {J.}~\bibnamefont {Schneeloch}}, \emph {et~al.},\ }\bibfield  {title} {\enquote {\bibinfo {title} {Mapping the orbital wavefunction of the surface states in three-dimensional topological insulators},}\ }\href@noop {} {\bibfield  {journal} {\bibinfo  {journal} {Nat. Phys.}\ }\textbf {\bibinfo {volume} {9}},\ \bibinfo {pages} {499} (\bibinfo {year} {2013})}\BibitemShut {NoStop}%
\bibitem [{\citenamefont {Hsieh}\ \emph {et~al.}(2009)\citenamefont {Hsieh}, \citenamefont {Xia}, \citenamefont {Qian}, \citenamefont {Wray}, \citenamefont {Dil}, \citenamefont {Meier}, \citenamefont {Osterwalder}, \citenamefont {Patthey}, \citenamefont {Checkelsky}, \citenamefont {Ong} \emph {et~al.}}]{hsieh2009tunable}%
  \BibitemOpen
  \bibfield  {author} {\bibinfo {author} {\bibfnamefont {D.}~\bibnamefont {Hsieh}}, \bibinfo {author} {\bibfnamefont {Y.}~\bibnamefont {Xia}}, \bibinfo {author} {\bibfnamefont {D.}~\bibnamefont {Qian}}, \bibinfo {author} {\bibfnamefont {L.}~\bibnamefont {Wray}}, \bibinfo {author} {\bibfnamefont {J.}~\bibnamefont {Dil}}, \bibinfo {author} {\bibfnamefont {F.}~\bibnamefont {Meier}}, \bibinfo {author} {\bibfnamefont {J.}~\bibnamefont {Osterwalder}}, \bibinfo {author} {\bibfnamefont {L.}~\bibnamefont {Patthey}}, \bibinfo {author} {\bibfnamefont {J.}~\bibnamefont {Checkelsky}}, \bibinfo {author} {\bibfnamefont {N.~P.}\ \bibnamefont {Ong}}, \emph {et~al.},\ }\bibfield  {title} {\enquote {\bibinfo {title} {A tunable topological insulator in the spin helical {D}irac transport regime},}\ }\href@noop {} {\bibfield  {journal} {\bibinfo  {journal} {Nature}\ }\textbf {\bibinfo {volume} {460}},\ \bibinfo {pages} {1101} (\bibinfo {year} {2009})}\BibitemShut {NoStop}%
\bibitem [{\citenamefont {Park}\ \emph {et~al.}(2012)\citenamefont {Park}, \citenamefont {Han}, \citenamefont {Kim}, \citenamefont {Koh}, \citenamefont {Kim}, \citenamefont {Lee}, \citenamefont {Choi}, \citenamefont {Han}, \citenamefont {Lee}, \citenamefont {Hur}, \citenamefont {Arita}, \citenamefont {Shimada}, \citenamefont {Namatame},\ and\ \citenamefont {Taniguchi}}]{Park2012chiralorbitalPRL}%
  \BibitemOpen
  \bibfield  {author} {\bibinfo {author} {\bibfnamefont {S.~R.}\ \bibnamefont {Park}}, \bibinfo {author} {\bibfnamefont {J.}~\bibnamefont {Han}}, \bibinfo {author} {\bibfnamefont {C.}~\bibnamefont {Kim}}, \bibinfo {author} {\bibfnamefont {Y.~Y.}\ \bibnamefont {Koh}}, \bibinfo {author} {\bibfnamefont {C.}~\bibnamefont {Kim}}, \bibinfo {author} {\bibfnamefont {H.}~\bibnamefont {Lee}}, \bibinfo {author} {\bibfnamefont {H.~J.}\ \bibnamefont {Choi}}, \bibinfo {author} {\bibfnamefont {J.~H.}\ \bibnamefont {Han}}, \bibinfo {author} {\bibfnamefont {K.~D.}\ \bibnamefont {Lee}}, \bibinfo {author} {\bibfnamefont {N.~J.}\ \bibnamefont {Hur}}, \bibinfo {author} {\bibfnamefont {M.}~\bibnamefont {Arita}}, \bibinfo {author} {\bibfnamefont {K.}~\bibnamefont {Shimada}}, \bibinfo {author} {\bibfnamefont {H.}~\bibnamefont {Namatame}},\ and\ \bibinfo {author} {\bibfnamefont {M.}~\bibnamefont {Taniguchi}},\ }\bibfield  {title} {\enquote {\bibinfo {title} {{Chiral Orbital-Angular Momentum in the Surface States of
  ${\mathrm{Bi}}_{2}{\mathrm{Se}}_{3}$}},}\ }\href {https://doi.org/10.1103/PhysRevLett.108.046805} {\bibfield  {journal} {\bibinfo  {journal} {Phys. Rev. Lett.}\ }\textbf {\bibinfo {volume} {108}},\ \bibinfo {pages} {046805} (\bibinfo {year} {2012})}\BibitemShut {NoStop}%
\bibitem [{\citenamefont {Kim}\ \emph {et~al.}(2014)\citenamefont {Kim}, \citenamefont {Yoshizawa}, \citenamefont {Ishida}, \citenamefont {Eto}, \citenamefont {Segawa}, \citenamefont {Ando}, \citenamefont {Shin},\ and\ \citenamefont {Komori}}]{Sunghun2014robustprotectionPRL}%
  \BibitemOpen
  \bibfield  {author} {\bibinfo {author} {\bibfnamefont {S.}~\bibnamefont {Kim}}, \bibinfo {author} {\bibfnamefont {S.}~\bibnamefont {Yoshizawa}}, \bibinfo {author} {\bibfnamefont {Y.}~\bibnamefont {Ishida}}, \bibinfo {author} {\bibfnamefont {K.}~\bibnamefont {Eto}}, \bibinfo {author} {\bibfnamefont {K.}~\bibnamefont {Segawa}}, \bibinfo {author} {\bibfnamefont {Y.}~\bibnamefont {Ando}}, \bibinfo {author} {\bibfnamefont {S.}~\bibnamefont {Shin}},\ and\ \bibinfo {author} {\bibfnamefont {F.}~\bibnamefont {Komori}},\ }\bibfield  {title} {\enquote {\bibinfo {title} {Robust protection from backscattering in the topological insulator ${\mathrm{bi}}_{1.5}{\mathrm{sb}}_{0.5}{\mathrm{te}}_{1.7}{\mathrm{se}}_{1.3}$},}\ }\href {https://doi.org/10.1103/PhysRevLett.112.136802} {\bibfield  {journal} {\bibinfo  {journal} {Phys. Rev. Lett.}\ }\textbf {\bibinfo {volume} {112}},\ \bibinfo {pages} {136802} (\bibinfo {year} {2014})}\BibitemShut {NoStop}%
\bibitem [{\citenamefont {Roushan}\ \emph {et~al.}(2009)\citenamefont {Roushan}, \citenamefont {Seo}, \citenamefont {Parker}, \citenamefont {Hor}, \citenamefont {Hsieh}, \citenamefont {Qian}, \citenamefont {Richardella}, \citenamefont {Hasan}, \citenamefont {Cava},\ and\ \citenamefont {Yazdani}}]{roushan2009topological}%
  \BibitemOpen
  \bibfield  {author} {\bibinfo {author} {\bibfnamefont {P.}~\bibnamefont {Roushan}}, \bibinfo {author} {\bibfnamefont {J.}~\bibnamefont {Seo}}, \bibinfo {author} {\bibfnamefont {C.~V.}\ \bibnamefont {Parker}}, \bibinfo {author} {\bibfnamefont {Y.~S.}\ \bibnamefont {Hor}}, \bibinfo {author} {\bibfnamefont {D.}~\bibnamefont {Hsieh}}, \bibinfo {author} {\bibfnamefont {D.}~\bibnamefont {Qian}}, \bibinfo {author} {\bibfnamefont {A.}~\bibnamefont {Richardella}}, \bibinfo {author} {\bibfnamefont {M.~Z.}\ \bibnamefont {Hasan}}, \bibinfo {author} {\bibfnamefont {R.~J.}\ \bibnamefont {Cava}},\ and\ \bibinfo {author} {\bibfnamefont {A.}~\bibnamefont {Yazdani}},\ }\bibfield  {title} {\enquote {\bibinfo {title} {Topological surface states protected from backscattering by chiral spin texture},}\ }\href@noop {} {\bibfield  {journal} {\bibinfo  {journal} {Nature}\ }\textbf {\bibinfo {volume} {460}},\ \bibinfo {pages} {1106} (\bibinfo {year} {2009})}\BibitemShut {NoStop}%
\bibitem [{\citenamefont {McIver}\ \emph {et~al.}(2012)\citenamefont {McIver}, \citenamefont {Hsieh}, \citenamefont {Steinberg}, \citenamefont {Jarillo-Herrero},\ and\ \citenamefont {Gedik}}]{mciver2012controlNNanotech}%
  \BibitemOpen
  \bibfield  {author} {\bibinfo {author} {\bibfnamefont {J.}~\bibnamefont {McIver}}, \bibinfo {author} {\bibfnamefont {D.}~\bibnamefont {Hsieh}}, \bibinfo {author} {\bibfnamefont {H.}~\bibnamefont {Steinberg}}, \bibinfo {author} {\bibfnamefont {P.}~\bibnamefont {Jarillo-Herrero}},\ and\ \bibinfo {author} {\bibfnamefont {N.}~\bibnamefont {Gedik}},\ }\bibfield  {title} {\enquote {\bibinfo {title} {{Control over topological insulator photocurrents with light polarization}},}\ }\href@noop {} {\bibfield  {journal} {\bibinfo  {journal} {Nat. Nanotechnol.}\ }\textbf {\bibinfo {volume} {7}},\ \bibinfo {pages} {96} (\bibinfo {year} {2012})}\BibitemShut {NoStop}%
\bibitem [{\citenamefont {Kastl}\ \emph {et~al.}(2015)\citenamefont {Kastl}, \citenamefont {Karnetzky}, \citenamefont {Karl},\ and\ \citenamefont {Holleitner}}]{kastl2015ultrafast}%
  \BibitemOpen
  \bibfield  {author} {\bibinfo {author} {\bibfnamefont {C.}~\bibnamefont {Kastl}}, \bibinfo {author} {\bibfnamefont {C.}~\bibnamefont {Karnetzky}}, \bibinfo {author} {\bibfnamefont {H.}~\bibnamefont {Karl}},\ and\ \bibinfo {author} {\bibfnamefont {A.~W.}\ \bibnamefont {Holleitner}},\ }\bibfield  {title} {\enquote {\bibinfo {title} {Ultrafast helicity control of surface currents in topological insulators with near-unity fidelity},}\ }\href@noop {} {\bibfield  {journal} {\bibinfo  {journal} {Nat. Commun.}\ }\textbf {\bibinfo {volume} {6}},\ \bibinfo {pages} {6617} (\bibinfo {year} {2015})}\BibitemShut {NoStop}%
\bibitem [{\citenamefont {Braun}\ \emph {et~al.}(2016)\citenamefont {Braun}, \citenamefont {Mussler}, \citenamefont {Hruban}, \citenamefont {Konczykowski}, \citenamefont {Schumann}, \citenamefont {Wolf}, \citenamefont {M{\"u}nzenberg}, \citenamefont {Perfetti},\ and\ \citenamefont {Kampfrath}}]{braun2016ultrafast}%
  \BibitemOpen
  \bibfield  {author} {\bibinfo {author} {\bibfnamefont {L.}~\bibnamefont {Braun}}, \bibinfo {author} {\bibfnamefont {G.}~\bibnamefont {Mussler}}, \bibinfo {author} {\bibfnamefont {A.}~\bibnamefont {Hruban}}, \bibinfo {author} {\bibfnamefont {M.}~\bibnamefont {Konczykowski}}, \bibinfo {author} {\bibfnamefont {T.}~\bibnamefont {Schumann}}, \bibinfo {author} {\bibfnamefont {M.}~\bibnamefont {Wolf}}, \bibinfo {author} {\bibfnamefont {M.}~\bibnamefont {M{\"u}nzenberg}}, \bibinfo {author} {\bibfnamefont {L.}~\bibnamefont {Perfetti}},\ and\ \bibinfo {author} {\bibfnamefont {T.}~\bibnamefont {Kampfrath}},\ }\bibfield  {title} {\enquote {\bibinfo {title} {Ultrafast photocurrents at the surface of the three-dimensional topological insulator $\mathrm{Bi_2Se_3}$},}\ }\href@noop {} {\bibfield  {journal} {\bibinfo  {journal} {Nat. Commun.}\ }\textbf {\bibinfo {volume} {7}},\ \bibinfo {pages} {13259} (\bibinfo {year} {2016})}\BibitemShut {NoStop}%
\bibitem [{\citenamefont {Kirilyuk}, \citenamefont {Kimel},\ and\ \citenamefont {Rasing}(2010)}]{manipulation_revmodphys}%
  \BibitemOpen
  \bibfield  {author} {\bibinfo {author} {\bibfnamefont {A.}~\bibnamefont {Kirilyuk}}, \bibinfo {author} {\bibfnamefont {A.~V.}\ \bibnamefont {Kimel}},\ and\ \bibinfo {author} {\bibfnamefont {T.}~\bibnamefont {Rasing}},\ }\bibfield  {title} {\enquote {\bibinfo {title} {Ultrafast optical manipulation of magnetic order},}\ }\href@noop {} {\bibfield  {journal} {\bibinfo  {journal} {Rev. Mod. Phys}\ }\textbf {\bibinfo {volume} {82}},\ \bibinfo {pages} {2731} (\bibinfo {year} {2010})}\BibitemShut {NoStop}%
\bibitem [{\citenamefont {Glazov}(2012)}]{coherentspin_review}%
  \BibitemOpen
  \bibfield  {author} {\bibinfo {author} {\bibfnamefont {M.}~\bibnamefont {Glazov}},\ }\bibfield  {title} {\enquote {\bibinfo {title} {Coherent spin dynamics of electrons and excitons in nanostructures (a review)},}\ }\href@noop {} {\bibfield  {journal} {\bibinfo  {journal} {Phys. Solid State}\ }\textbf {\bibinfo {volume} {54}},\ \bibinfo {pages} {1} (\bibinfo {year} {2012})}\BibitemShut {NoStop}%
\bibitem [{\citenamefont {Kuroda}\ \emph {et~al.}(2017)\citenamefont {Kuroda}, \citenamefont {Reimann}, \citenamefont {Kokh}, \citenamefont {Tereshchenko}, \citenamefont {Kimura}, \citenamefont {G{\"u}dde},\ and\ \citenamefont {H{\"o}fer}}]{kuroda2017ultrafastPRB}%
  \BibitemOpen
  \bibfield  {author} {\bibinfo {author} {\bibfnamefont {K.}~\bibnamefont {Kuroda}}, \bibinfo {author} {\bibfnamefont {J.}~\bibnamefont {Reimann}}, \bibinfo {author} {\bibfnamefont {K.}~\bibnamefont {Kokh}}, \bibinfo {author} {\bibfnamefont {O.}~\bibnamefont {Tereshchenko}}, \bibinfo {author} {\bibfnamefont {A.}~\bibnamefont {Kimura}}, \bibinfo {author} {\bibfnamefont {J.}~\bibnamefont {G{\"u}dde}},\ and\ \bibinfo {author} {\bibfnamefont {U.}~\bibnamefont {H{\"o}fer}},\ }\bibfield  {title} {\enquote {\bibinfo {title} {Ultrafast energy-and momentum-resolved surface {D}irac photocurrents in the topological insulator $\mathrm{Sb_{2}Te_{3}}$},}\ }\href@noop {} {\bibfield  {journal} {\bibinfo  {journal} {Phys. Rev. B}\ }\textbf {\bibinfo {volume} {95}},\ \bibinfo {pages} {081103} (\bibinfo {year} {2017})}\BibitemShut {NoStop}%
\bibitem [{\citenamefont {Kuroda}\ \emph {et~al.}(2016)\citenamefont {Kuroda}, \citenamefont {Reimann}, \citenamefont {G{\"u}dde},\ and\ \citenamefont {H{\"o}fer}}]{kuroda2016generationPRL}%
  \BibitemOpen
  \bibfield  {author} {\bibinfo {author} {\bibfnamefont {K.}~\bibnamefont {Kuroda}}, \bibinfo {author} {\bibfnamefont {J.}~\bibnamefont {Reimann}}, \bibinfo {author} {\bibfnamefont {J.}~\bibnamefont {G{\"u}dde}},\ and\ \bibinfo {author} {\bibfnamefont {U.}~\bibnamefont {H{\"o}fer}},\ }\bibfield  {title} {\enquote {\bibinfo {title} {{Generation of Transient Photocurrents in the Topological Surface State of $\mathrm{Sb_{2}Te_{3}}$ by Direct Optical Excitation with Midinfrared Pulses}},}\ }\href@noop {} {\bibfield  {journal} {\bibinfo  {journal} {Phys. Rev. Lett.}\ }\textbf {\bibinfo {volume} {116}},\ \bibinfo {pages} {076801} (\bibinfo {year} {2016})}\BibitemShut {NoStop}%
\bibitem [{\citenamefont {Soifer}\ \emph {et~al.}(2019)\citenamefont {Soifer}, \citenamefont {Gauthier}, \citenamefont {Kemper}, \citenamefont {Rotundu}, \citenamefont {Yang}, \citenamefont {Xiong}, \citenamefont {Lu}, \citenamefont {Hashimoto}, \citenamefont {Kirchmann}, \citenamefont {Sobota} \emph {et~al.}}]{soifer2019bandresolvedPRL}%
  \BibitemOpen
  \bibfield  {author} {\bibinfo {author} {\bibfnamefont {H.}~\bibnamefont {Soifer}}, \bibinfo {author} {\bibfnamefont {A.}~\bibnamefont {Gauthier}}, \bibinfo {author} {\bibfnamefont {A.~F.}\ \bibnamefont {Kemper}}, \bibinfo {author} {\bibfnamefont {C.~R.}\ \bibnamefont {Rotundu}}, \bibinfo {author} {\bibfnamefont {S.-L.}\ \bibnamefont {Yang}}, \bibinfo {author} {\bibfnamefont {H.}~\bibnamefont {Xiong}}, \bibinfo {author} {\bibfnamefont {D.}~\bibnamefont {Lu}}, \bibinfo {author} {\bibfnamefont {M.}~\bibnamefont {Hashimoto}}, \bibinfo {author} {\bibfnamefont {P.~S.}\ \bibnamefont {Kirchmann}}, \bibinfo {author} {\bibfnamefont {J.~A.}\ \bibnamefont {Sobota}}, \emph {et~al.},\ }\bibfield  {title} {\enquote {\bibinfo {title} {{Band-Resolved Imaging of Photocurrent in a Topological Insulator}},}\ }\href@noop {} {\bibfield  {journal} {\bibinfo  {journal} {Phys. Rev. Lett.}\ }\textbf {\bibinfo {volume} {122}},\ \bibinfo {pages} {167401} (\bibinfo {year} {2019})}\BibitemShut {NoStop}%
\bibitem [{\citenamefont {Ketterl}\ \emph {et~al.}(2018)\citenamefont {Ketterl}, \citenamefont {Otto}, \citenamefont {Bastian}, \citenamefont {Andres}, \citenamefont {Gahl}, \citenamefont {Min{\'a}r}, \citenamefont {Ebert}, \citenamefont {Braun}, \citenamefont {Tereshchenko}, \citenamefont {Kokh} \emph {et~al.}}]{ketterl2018origin}%
  \BibitemOpen
  \bibfield  {author} {\bibinfo {author} {\bibfnamefont {A.}~\bibnamefont {Ketterl}}, \bibinfo {author} {\bibfnamefont {S.}~\bibnamefont {Otto}}, \bibinfo {author} {\bibfnamefont {M.}~\bibnamefont {Bastian}}, \bibinfo {author} {\bibfnamefont {B.}~\bibnamefont {Andres}}, \bibinfo {author} {\bibfnamefont {C.}~\bibnamefont {Gahl}}, \bibinfo {author} {\bibfnamefont {J.}~\bibnamefont {Min{\'a}r}}, \bibinfo {author} {\bibfnamefont {H.}~\bibnamefont {Ebert}}, \bibinfo {author} {\bibfnamefont {J.}~\bibnamefont {Braun}}, \bibinfo {author} {\bibfnamefont {O.~E.}\ \bibnamefont {Tereshchenko}}, \bibinfo {author} {\bibfnamefont {K.~A.}\ \bibnamefont {Kokh}}, \emph {et~al.},\ }\bibfield  {title} {\enquote {\bibinfo {title} {Origin of spin-polarized photocurrents in the topological surface states of $\mathrm{Bi_2Se_3}$},}\ }\href@noop {} {\bibfield  {journal} {\bibinfo  {journal} {Phys. Rev. B}\ }\textbf {\bibinfo {volume} {98}},\ \bibinfo {pages} {155406} (\bibinfo {year} {2018})}\BibitemShut {NoStop}%
\bibitem [{\citenamefont {Reimann}\ \emph {et~al.}(2014)\citenamefont {Reimann}, \citenamefont {G{\"u}dde}, \citenamefont {Kuroda}, \citenamefont {Chulkov},\ and\ \citenamefont {H{\"o}fer}}]{reimann2014spectroscopy}%
  \BibitemOpen
  \bibfield  {author} {\bibinfo {author} {\bibfnamefont {J.}~\bibnamefont {Reimann}}, \bibinfo {author} {\bibfnamefont {J.}~\bibnamefont {G{\"u}dde}}, \bibinfo {author} {\bibfnamefont {K.}~\bibnamefont {Kuroda}}, \bibinfo {author} {\bibfnamefont {E.~V.}\ \bibnamefont {Chulkov}},\ and\ \bibinfo {author} {\bibfnamefont {U.}~\bibnamefont {H{\"o}fer}},\ }\bibfield  {title} {\enquote {\bibinfo {title} {Spectroscopy and dynamics of unoccupied electronic states of the topological insulators $\mathrm{Sb_2Te_3}$ and $\mathrm{Sb_2Te_2S}$},}\ }\href@noop {} {\bibfield  {journal} {\bibinfo  {journal} {Phys. Rev. B}\ }\textbf {\bibinfo {volume} {90}},\ \bibinfo {pages} {081106} (\bibinfo {year} {2014})}\BibitemShut {NoStop}%
\bibitem [{\citenamefont {Reimann}\ \emph {et~al.}(2018)\citenamefont {Reimann}, \citenamefont {Schlauderer}, \citenamefont {Schmid}, \citenamefont {Langer}, \citenamefont {Baierl}, \citenamefont {Kokh}, \citenamefont {Tereshchenko}, \citenamefont {Kimura}, \citenamefont {Lange}, \citenamefont {G{\"u}dde} \emph {et~al.}}]{RHuber2018subcycleNature}%
  \BibitemOpen
  \bibfield  {author} {\bibinfo {author} {\bibfnamefont {J.}~\bibnamefont {Reimann}}, \bibinfo {author} {\bibfnamefont {S.}~\bibnamefont {Schlauderer}}, \bibinfo {author} {\bibfnamefont {C.}~\bibnamefont {Schmid}}, \bibinfo {author} {\bibfnamefont {F.}~\bibnamefont {Langer}}, \bibinfo {author} {\bibfnamefont {S.}~\bibnamefont {Baierl}}, \bibinfo {author} {\bibfnamefont {K.}~\bibnamefont {Kokh}}, \bibinfo {author} {\bibfnamefont {O.}~\bibnamefont {Tereshchenko}}, \bibinfo {author} {\bibfnamefont {A.}~\bibnamefont {Kimura}}, \bibinfo {author} {\bibfnamefont {C.}~\bibnamefont {Lange}}, \bibinfo {author} {\bibfnamefont {J.}~\bibnamefont {G{\"u}dde}}, \emph {et~al.},\ }\bibfield  {title} {\enquote {\bibinfo {title} {Subcycle observation of lightwave-driven {D}irac currents in a topological surface band},}\ }\href@noop {} {\bibfield  {journal} {\bibinfo  {journal} {Nature}\ }\textbf {\bibinfo {volume} {562}},\ \bibinfo {pages} {396} (\bibinfo {year} {2018})}\BibitemShut {NoStop}%
\bibitem [{\citenamefont {Hsieh}\ \emph {et~al.}(2011)\citenamefont {Hsieh}, \citenamefont {Mahmood}, \citenamefont {McIver}, \citenamefont {Gardner}, \citenamefont {Lee},\ and\ \citenamefont {Gedik}}]{DHsiehPRLselective}%
  \BibitemOpen
  \bibfield  {author} {\bibinfo {author} {\bibfnamefont {D.}~\bibnamefont {Hsieh}}, \bibinfo {author} {\bibfnamefont {F.}~\bibnamefont {Mahmood}}, \bibinfo {author} {\bibfnamefont {J.~W.}\ \bibnamefont {McIver}}, \bibinfo {author} {\bibfnamefont {D.~R.}\ \bibnamefont {Gardner}}, \bibinfo {author} {\bibfnamefont {Y.~S.}\ \bibnamefont {Lee}},\ and\ \bibinfo {author} {\bibfnamefont {N.}~\bibnamefont {Gedik}},\ }\bibfield  {title} {\enquote {\bibinfo {title} {{Selective Probing of Photoinduced Charge and Spin Dynamics in the Bulk and Surface of a Topological Insulator}},}\ }\href {https://doi.org/10.1103/PhysRevLett.107.077401} {\bibfield  {journal} {\bibinfo  {journal} {Phys. Rev. Lett.}\ }\textbf {\bibinfo {volume} {107}},\ \bibinfo {pages} {077401} (\bibinfo {year} {2011})}\BibitemShut {NoStop}%
\bibitem [{\citenamefont {Wang}\ \emph {et~al.}(2016)\citenamefont {Wang}, \citenamefont {Qiao}, \citenamefont {Jiang}, \citenamefont {Luo},\ and\ \citenamefont {Qi}}]{MCwangPRLunraveling}%
  \BibitemOpen
  \bibfield  {author} {\bibinfo {author} {\bibfnamefont {M.~C.}\ \bibnamefont {Wang}}, \bibinfo {author} {\bibfnamefont {S.}~\bibnamefont {Qiao}}, \bibinfo {author} {\bibfnamefont {Z.}~\bibnamefont {Jiang}}, \bibinfo {author} {\bibfnamefont {S.~N.}\ \bibnamefont {Luo}},\ and\ \bibinfo {author} {\bibfnamefont {J.}~\bibnamefont {Qi}},\ }\bibfield  {title} {\enquote {\bibinfo {title} {{Unraveling Photoinduced Spin Dynamics in the Topological Insulator ${\mathrm{Bi}}_{2}{\mathrm{Se}}_{3}$}},}\ }\href {https://doi.org/10.1103/PhysRevLett.116.036601} {\bibfield  {journal} {\bibinfo  {journal} {Phys. Rev. Lett.}\ }\textbf {\bibinfo {volume} {116}},\ \bibinfo {pages} {036601} (\bibinfo {year} {2016})}\BibitemShut {NoStop}%
\bibitem [{\citenamefont {Boschini}\ \emph {et~al.}(2015)\citenamefont {Boschini}, \citenamefont {Mansurova}, \citenamefont {Mussler}, \citenamefont {Kampmeier}, \citenamefont {Gr{\"u}tzmacher}, \citenamefont {Braun}, \citenamefont {Katmis}, \citenamefont {Moodera}, \citenamefont {Dallera}, \citenamefont {Carpene} \emph {et~al.}}]{boschini2015SciRep}%
  \BibitemOpen
  \bibfield  {author} {\bibinfo {author} {\bibfnamefont {F.}~\bibnamefont {Boschini}}, \bibinfo {author} {\bibfnamefont {M.}~\bibnamefont {Mansurova}}, \bibinfo {author} {\bibfnamefont {G.}~\bibnamefont {Mussler}}, \bibinfo {author} {\bibfnamefont {J.}~\bibnamefont {Kampmeier}}, \bibinfo {author} {\bibfnamefont {D.}~\bibnamefont {Gr{\"u}tzmacher}}, \bibinfo {author} {\bibfnamefont {L.}~\bibnamefont {Braun}}, \bibinfo {author} {\bibfnamefont {F.}~\bibnamefont {Katmis}}, \bibinfo {author} {\bibfnamefont {J.~S.}\ \bibnamefont {Moodera}}, \bibinfo {author} {\bibfnamefont {C.}~\bibnamefont {Dallera}}, \bibinfo {author} {\bibfnamefont {E.}~\bibnamefont {Carpene}}, \emph {et~al.},\ }\bibfield  {title} {\enquote {\bibinfo {title} {Coherent ultrafast spin-dynamics probed in three dimensional topological insulators},}\ }\href@noop {} {\bibfield  {journal} {\bibinfo  {journal} {Sci. Rep.}\ }\textbf {\bibinfo {volume} {5}},\ \bibinfo {pages} {15304} (\bibinfo {year} {2015})}\BibitemShut {NoStop}%
\bibitem [{\citenamefont {Mondal}\ \emph {et~al.}(2018)\citenamefont {Mondal}, \citenamefont {Aihara}, \citenamefont {Saito}, \citenamefont {Fons}, \citenamefont {Kolobov}, \citenamefont {Tominaga},\ and\ \citenamefont {Hase}}]{mondal2018topological}%
  \BibitemOpen
  \bibfield  {author} {\bibinfo {author} {\bibfnamefont {R.}~\bibnamefont {Mondal}}, \bibinfo {author} {\bibfnamefont {Y.}~\bibnamefont {Aihara}}, \bibinfo {author} {\bibfnamefont {Y.}~\bibnamefont {Saito}}, \bibinfo {author} {\bibfnamefont {P.}~\bibnamefont {Fons}}, \bibinfo {author} {\bibfnamefont {A.~V.}\ \bibnamefont {Kolobov}}, \bibinfo {author} {\bibfnamefont {J.}~\bibnamefont {Tominaga}},\ and\ \bibinfo {author} {\bibfnamefont {M.}~\bibnamefont {Hase}},\ }\bibfield  {title} {\enquote {\bibinfo {title} {{Topological Phase Buried in a Chalcogenide Superlattice Monitored by Helicity-Dependent Kerr Measurement}},}\ }\href@noop {} {\bibfield  {journal} {\bibinfo  {journal} {ACS Appl. Mater. Interfaces}\ }\textbf {\bibinfo {volume} {10}},\ \bibinfo {pages} {26781} (\bibinfo {year} {2018})}\BibitemShut {NoStop}%
\bibitem [{\citenamefont {Kimel}\ \emph {et~al.}(2005)\citenamefont {Kimel}, \citenamefont {Kirilyuk}, \citenamefont {Usachev}, \citenamefont {Pisarev}, \citenamefont {Balbashov},\ and\ \citenamefont {Rasing}}]{Kimel2005Nature}%
  \BibitemOpen
  \bibfield  {author} {\bibinfo {author} {\bibfnamefont {A.~V.}\ \bibnamefont {Kimel}}, \bibinfo {author} {\bibfnamefont {A.}~\bibnamefont {Kirilyuk}}, \bibinfo {author} {\bibfnamefont {P.~A.}\ \bibnamefont {Usachev}}, \bibinfo {author} {\bibfnamefont {R.~V.}\ \bibnamefont {Pisarev}}, \bibinfo {author} {\bibfnamefont {A.~M.}\ \bibnamefont {Balbashov}},\ and\ \bibinfo {author} {\bibfnamefont {T.}~\bibnamefont {Rasing}},\ }\bibfield  {title} {\enquote {\bibinfo {title} {Ultrafast non-thermal control of magnetization by instantaneous photomagnetic pulses},}\ }\href@noop {} {\bibfield  {journal} {\bibinfo  {journal} {Nature}\ }\textbf {\bibinfo {volume} {435}},\ \bibinfo {pages} {655} (\bibinfo {year} {2005})}\BibitemShut {NoStop}%
\bibitem [{\citenamefont {Kini}\ \emph {et~al.}(2008)\citenamefont {Kini}, \citenamefont {Nontapot}, \citenamefont {Khodaparast}, \citenamefont {Welser},\ and\ \citenamefont {Guido}}]{JAPInAsKini2008}%
  \BibitemOpen
  \bibfield  {author} {\bibinfo {author} {\bibfnamefont {R.~N.}\ \bibnamefont {Kini}}, \bibinfo {author} {\bibfnamefont {K.}~\bibnamefont {Nontapot}}, \bibinfo {author} {\bibfnamefont {G.~A.}\ \bibnamefont {Khodaparast}}, \bibinfo {author} {\bibfnamefont {R.~E.}\ \bibnamefont {Welser}},\ and\ \bibinfo {author} {\bibfnamefont {L.~J.}\ \bibnamefont {Guido}},\ }\bibfield  {title} {\enquote {\bibinfo {title} {Time resolved measurements of spin and carrier dynamics in {InAs} films},}\ }\href@noop {} {\bibfield  {journal} {\bibinfo  {journal} {J. Appl. Phys.}\ }\textbf {\bibinfo {volume} {103}} (\bibinfo {year} {2008})}\BibitemShut {NoStop}%
\bibitem [{\citenamefont {Zhao}\ \emph {et~al.}(2020)\citenamefont {Zhao}, \citenamefont {Su}, \citenamefont {Huang}, \citenamefont {Wu}, \citenamefont {Fong}, \citenamefont {Feng},\ and\ \citenamefont {Xiong}}]{NatComPerovskite2020}%
  \BibitemOpen
  \bibfield  {author} {\bibinfo {author} {\bibfnamefont {W.}~\bibnamefont {Zhao}}, \bibinfo {author} {\bibfnamefont {R.}~\bibnamefont {Su}}, \bibinfo {author} {\bibfnamefont {Y.}~\bibnamefont {Huang}}, \bibinfo {author} {\bibfnamefont {J.}~\bibnamefont {Wu}}, \bibinfo {author} {\bibfnamefont {C.~F.}\ \bibnamefont {Fong}}, \bibinfo {author} {\bibfnamefont {J.}~\bibnamefont {Feng}},\ and\ \bibinfo {author} {\bibfnamefont {Q.}~\bibnamefont {Xiong}},\ }\bibfield  {title} {\enquote {\bibinfo {title} {Transient circular dichroism and exciton spin dynamics in all-inorganic halide perovskites},}\ }\href@noop {} {\bibfield  {journal} {\bibinfo  {journal} {Nat. Commun.}\ }\textbf {\bibinfo {volume} {11}},\ \bibinfo {pages} {5665} (\bibinfo {year} {2020})}\BibitemShut {NoStop}%
\bibitem [{\citenamefont {Kimel}\ \emph {et~al.}(2001)\citenamefont {Kimel}, \citenamefont {Bentivegna}, \citenamefont {Gridnev}, \citenamefont {Pavlov}, \citenamefont {Pisarev},\ and\ \citenamefont {Rasing}}]{kimel2001roomtemp}%
  \BibitemOpen
  \bibfield  {author} {\bibinfo {author} {\bibfnamefont {A.}~\bibnamefont {Kimel}}, \bibinfo {author} {\bibfnamefont {F.}~\bibnamefont {Bentivegna}}, \bibinfo {author} {\bibfnamefont {V.}~\bibnamefont {Gridnev}}, \bibinfo {author} {\bibfnamefont {V.}~\bibnamefont {Pavlov}}, \bibinfo {author} {\bibfnamefont {R.}~\bibnamefont {Pisarev}},\ and\ \bibinfo {author} {\bibfnamefont {T.}~\bibnamefont {Rasing}},\ }\bibfield  {title} {\enquote {\bibinfo {title} {Room-temperature ultrafast carrier and spin dynamics in {GaAs} probed by the photoinduced magneto-optical {K}err effect},}\ }\href@noop {} {\bibfield  {journal} {\bibinfo  {journal} {Phys. Rev. B}\ }\textbf {\bibinfo {volume} {63}},\ \bibinfo {pages} {235201} (\bibinfo {year} {2001})}\BibitemShut {NoStop}%
\bibitem [{\citenamefont {Zhu}\ \emph {et~al.}(2014)\citenamefont {Zhu}, \citenamefont {Zhang}, \citenamefont {Glazov}, \citenamefont {Urbaszek}, \citenamefont {Amand}, \citenamefont {Ji}, \citenamefont {Liu},\ and\ \citenamefont {Marie}}]{PRBWSeExciton2014}%
  \BibitemOpen
  \bibfield  {author} {\bibinfo {author} {\bibfnamefont {C.}~\bibnamefont {Zhu}}, \bibinfo {author} {\bibfnamefont {K.}~\bibnamefont {Zhang}}, \bibinfo {author} {\bibfnamefont {M.}~\bibnamefont {Glazov}}, \bibinfo {author} {\bibfnamefont {B.}~\bibnamefont {Urbaszek}}, \bibinfo {author} {\bibfnamefont {T.}~\bibnamefont {Amand}}, \bibinfo {author} {\bibfnamefont {Z.}~\bibnamefont {Ji}}, \bibinfo {author} {\bibfnamefont {B.}~\bibnamefont {Liu}},\ and\ \bibinfo {author} {\bibfnamefont {X.}~\bibnamefont {Marie}},\ }\bibfield  {title} {\enquote {\bibinfo {title} {Exciton valley dynamics probed by {K}err rotation in $\mathrm{WSe_2}$ monolayers},}\ }\href@noop {} {\bibfield  {journal} {\bibinfo  {journal} {Phys. Rev. B}\ }\textbf {\bibinfo {volume} {90}},\ \bibinfo {pages} {161302} (\bibinfo {year} {2014})}\BibitemShut {NoStop}%
\bibitem [{\citenamefont {Chernikov}\ \emph {et~al.}(2015{\natexlab{a}})\citenamefont {Chernikov}, \citenamefont {Ruppert}, \citenamefont {Hill}, \citenamefont {Rigosi},\ and\ \citenamefont {Heinz}}]{chernikov2015population}%
  \BibitemOpen
  \bibfield  {author} {\bibinfo {author} {\bibfnamefont {A.}~\bibnamefont {Chernikov}}, \bibinfo {author} {\bibfnamefont {C.}~\bibnamefont {Ruppert}}, \bibinfo {author} {\bibfnamefont {H.~M.}\ \bibnamefont {Hill}}, \bibinfo {author} {\bibfnamefont {A.~F.}\ \bibnamefont {Rigosi}},\ and\ \bibinfo {author} {\bibfnamefont {T.~F.}\ \bibnamefont {Heinz}},\ }\bibfield  {title} {\enquote {\bibinfo {title} {Population inversion and giant bandgap renormalization in atomically thin $\mathrm{WS_2}$ layers},}\ }\href@noop {} {\bibfield  {journal} {\bibinfo  {journal} {Nat. Photonics}\ }\textbf {\bibinfo {volume} {9}},\ \bibinfo {pages} {466--470} (\bibinfo {year} {2015}{\natexlab{a}})}\BibitemShut {NoStop}%
\bibitem [{\citenamefont {Fukuda}\ \emph {et~al.}(2023)\citenamefont {Fukuda}, \citenamefont {Ozaki}, \citenamefont {Jeong}, \citenamefont {Arashida}, \citenamefont {En-ya}, \citenamefont {Yoshida}, \citenamefont {Fons}, \citenamefont {Fujita}, \citenamefont {Ueno}, \citenamefont {Hase},\ and\ \citenamefont {Hada}}]{Fukuda2023HighDensity}%
  \BibitemOpen
  \bibfield  {author} {\bibinfo {author} {\bibfnamefont {T.}~\bibnamefont {Fukuda}}, \bibinfo {author} {\bibfnamefont {U.}~\bibnamefont {Ozaki}}, \bibinfo {author} {\bibfnamefont {S.}~\bibnamefont {Jeong}}, \bibinfo {author} {\bibfnamefont {Y.}~\bibnamefont {Arashida}}, \bibinfo {author} {\bibfnamefont {K.}~\bibnamefont {En-ya}}, \bibinfo {author} {\bibfnamefont {S.}~\bibnamefont {Yoshida}}, \bibinfo {author} {\bibfnamefont {P.~J.}\ \bibnamefont {Fons}}, \bibinfo {author} {\bibfnamefont {J.-i.}\ \bibnamefont {Fujita}}, \bibinfo {author} {\bibfnamefont {K.}~\bibnamefont {Ueno}}, \bibinfo {author} {\bibfnamefont {M.}~\bibnamefont {Hase}},\ and\ \bibinfo {author} {\bibfnamefont {M.}~\bibnamefont {Hada}},\ }\bibfield  {title} {\enquote {\bibinfo {title} {Photoinduced {S}tructural {D}ynamics of $\mathrm{2H-MoTe_2}$ {U}nder {E}xtremely {H}igh-{D}ensity {E}xcitation {C}onditions},}\ }\href {https://doi.org/10.1021/acs.jpcc.3c02838} {\bibfield  {journal} {\bibinfo  {journal} {J. Phys. Chem. C}\ }\textbf {\bibinfo
  {volume} {127}},\ \bibinfo {pages} {13149--13156} (\bibinfo {year} {2023})}\BibitemShut {NoStop}%
\bibitem [{\citenamefont {Iwasaki}\ \emph {et~al.}(2023)\citenamefont {Iwasaki}, \citenamefont {Fukuda}, \citenamefont {Noyama}, \citenamefont {Akei}, \citenamefont {Shigekawa}, \citenamefont {Fons}, \citenamefont {Hase}, \citenamefont {Arashida},\ and\ \citenamefont {Hada}}]{iwasakiFukuda2023APL}%
  \BibitemOpen
  \bibfield  {author} {\bibinfo {author} {\bibfnamefont {Y.}~\bibnamefont {Iwasaki}}, \bibinfo {author} {\bibfnamefont {T.}~\bibnamefont {Fukuda}}, \bibinfo {author} {\bibfnamefont {G.}~\bibnamefont {Noyama}}, \bibinfo {author} {\bibfnamefont {M.}~\bibnamefont {Akei}}, \bibinfo {author} {\bibfnamefont {H.}~\bibnamefont {Shigekawa}}, \bibinfo {author} {\bibfnamefont {P.~J.}\ \bibnamefont {Fons}}, \bibinfo {author} {\bibfnamefont {M.}~\bibnamefont {Hase}}, \bibinfo {author} {\bibfnamefont {Y.}~\bibnamefont {Arashida}},\ and\ \bibinfo {author} {\bibfnamefont {M.}~\bibnamefont {Hada}},\ }\bibfield  {title} {\enquote {\bibinfo {title} {Electronic intraband scattering in a transition-metal dichalcogenide observed by double-excitation ultrafast electron diffraction},}\ }\href@noop {} {\bibfield  {journal} {\bibinfo  {journal} {Appl. Phys. Lett.}\ }\textbf {\bibinfo {volume} {123}},\ \bibinfo {pages} {181901} (\bibinfo {year} {2023})}\BibitemShut {NoStop}%
\bibitem [{\citenamefont {Fukuda}\ \emph {et~al.}(2024)\citenamefont {Fukuda}, \citenamefont {Makino}, \citenamefont {Saito}, \citenamefont {Fons}, \citenamefont {Ando}, \citenamefont {Mori}, \citenamefont {Ishikawa}, \citenamefont {Ueno}, \citenamefont {Afalla},\ and\ \citenamefont {Hase}}]{fukuda2024coherent}%
  \BibitemOpen
  \bibfield  {author} {\bibinfo {author} {\bibfnamefont {T.}~\bibnamefont {Fukuda}}, \bibinfo {author} {\bibfnamefont {K.}~\bibnamefont {Makino}}, \bibinfo {author} {\bibfnamefont {Y.}~\bibnamefont {Saito}}, \bibinfo {author} {\bibfnamefont {P.}~\bibnamefont {Fons}}, \bibinfo {author} {\bibfnamefont {A.}~\bibnamefont {Ando}}, \bibinfo {author} {\bibfnamefont {T.}~\bibnamefont {Mori}}, \bibinfo {author} {\bibfnamefont {R.}~\bibnamefont {Ishikawa}}, \bibinfo {author} {\bibfnamefont {K.}~\bibnamefont {Ueno}}, \bibinfo {author} {\bibfnamefont {J.}~\bibnamefont {Afalla}},\ and\ \bibinfo {author} {\bibfnamefont {M.}~\bibnamefont {Hase}},\ }\bibfield  {title} {\enquote {\bibinfo {title} {Coherent optical response driven by non-equilibrium electron--phonon dynamics in a layered transition-metal dichalcogenide},}\ }\href@noop {} {\bibfield  {journal} {\bibinfo  {journal} {APL Mater.}\ }\textbf {\bibinfo {volume} {12}},\ \bibinfo {pages} {021102} (\bibinfo {year} {2024})}\BibitemShut {NoStop}%
\bibitem [{\citenamefont {Shrivastava}\ \emph {et~al.}(2022)\citenamefont {Shrivastava}, \citenamefont {Hazarika}, \citenamefont {Aneesh}, \citenamefont {Mandal}, \citenamefont {Beard},\ and\ \citenamefont {Adarsh}}]{PRBGiantSSBGR}%
  \BibitemOpen
  \bibfield  {author} {\bibinfo {author} {\bibfnamefont {M.}~\bibnamefont {Shrivastava}}, \bibinfo {author} {\bibfnamefont {A.}~\bibnamefont {Hazarika}}, \bibinfo {author} {\bibfnamefont {J.}~\bibnamefont {Aneesh}}, \bibinfo {author} {\bibfnamefont {D.}~\bibnamefont {Mandal}}, \bibinfo {author} {\bibfnamefont {M.~C.}\ \bibnamefont {Beard}},\ and\ \bibinfo {author} {\bibfnamefont {K.~V.}\ \bibnamefont {Adarsh}},\ }\bibfield  {title} {\enquote {\bibinfo {title} {Giant spin-selective bandgap renormalization in $\mathrm{CsPbBr}_{3}$ colloidal nanocrystals},}\ }\href {https://doi.org/10.1103/PhysRevB.106.L041404} {\bibfield  {journal} {\bibinfo  {journal} {Phys. Rev. B}\ }\textbf {\bibinfo {volume} {106}},\ \bibinfo {pages} {L041404} (\bibinfo {year} {2022})}\BibitemShut {NoStop}%
\bibitem [{\citenamefont {Nemec}\ \emph {et~al.}(2005)\citenamefont {Nemec}, \citenamefont {Kerachian}, \citenamefont {van Driel},\ and\ \citenamefont {Smirl}}]{PRBspindepGaAs}%
  \BibitemOpen
  \bibfield  {author} {\bibinfo {author} {\bibfnamefont {P.}~\bibnamefont {Nemec}}, \bibinfo {author} {\bibfnamefont {Y.}~\bibnamefont {Kerachian}}, \bibinfo {author} {\bibfnamefont {H.~M.}\ \bibnamefont {van Driel}},\ and\ \bibinfo {author} {\bibfnamefont {A.~L.}\ \bibnamefont {Smirl}},\ }\bibfield  {title} {\enquote {\bibinfo {title} {Spin-dependent electron many-body effects in {GaAs}},}\ }\href {https://doi.org/10.1103/PhysRevB.72.245202} {\bibfield  {journal} {\bibinfo  {journal} {Phys. Rev. B}\ }\textbf {\bibinfo {volume} {72}},\ \bibinfo {pages} {245202} (\bibinfo {year} {2005})}\BibitemShut {NoStop}%
\bibitem [{\citenamefont {Horodysk{\'a}}\ \emph {et~al.}(2014)\citenamefont {Horodysk{\'a}}, \citenamefont {N{\v{e}}mec}, \citenamefont {Novotn{\`y}}, \citenamefont {Troj{\'a}nek},\ and\ \citenamefont {Mal{\`y}}}]{JAP2014spindepCdTe}%
  \BibitemOpen
  \bibfield  {author} {\bibinfo {author} {\bibfnamefont {P.}~\bibnamefont {Horodysk{\'a}}}, \bibinfo {author} {\bibfnamefont {P.}~\bibnamefont {N{\v{e}}mec}}, \bibinfo {author} {\bibfnamefont {T.}~\bibnamefont {Novotn{\`y}}}, \bibinfo {author} {\bibfnamefont {F.}~\bibnamefont {Troj{\'a}nek}},\ and\ \bibinfo {author} {\bibfnamefont {P.}~\bibnamefont {Mal{\`y}}},\ }\bibfield  {title} {\enquote {\bibinfo {title} {Experimental observation of spin-dependent electron many-body effects in {CdTe}},}\ }\href@noop {} {\bibfield  {journal} {\bibinfo  {journal} {J. Appl. Phys.}\ }\textbf {\bibinfo {volume} {116}} (\bibinfo {year} {2014})}\BibitemShut {NoStop}%
\bibitem [{\citenamefont {Eddrief}, \citenamefont {Vidal},\ and\ \citenamefont {Gallas}(2016)}]{eddrief2016bulktoulthin}%
  \BibitemOpen
  \bibfield  {author} {\bibinfo {author} {\bibfnamefont {M.}~\bibnamefont {Eddrief}}, \bibinfo {author} {\bibfnamefont {F.}~\bibnamefont {Vidal}},\ and\ \bibinfo {author} {\bibfnamefont {B.}~\bibnamefont {Gallas}},\ }\bibfield  {title} {\enquote {\bibinfo {title} {Optical properties of $\mathrm{Bi_2Se_3}$: from bulk to ultrathin films},}\ }\href@noop {} {\bibfield  {journal} {\bibinfo  {journal} {J. Phys. D}\ }\textbf {\bibinfo {volume} {49}},\ \bibinfo {pages} {505304} (\bibinfo {year} {2016})}\BibitemShut {NoStop}%
\bibitem [{\citenamefont {Kresse}\ and\ \citenamefont {Hafner}(1993)}]{Kresse1993DFT}%
  \BibitemOpen
  \bibfield  {author} {\bibinfo {author} {\bibfnamefont {G.}~\bibnamefont {Kresse}}\ and\ \bibinfo {author} {\bibfnamefont {J.}~\bibnamefont {Hafner}},\ }\bibfield  {title} {\enquote {\bibinfo {title} {Ab initio molecular dynamics for liquid metals},}\ }\href {https://doi.org/10.1103/PhysRevB.47.558} {\bibfield  {journal} {\bibinfo  {journal} {Phys. Rev. B}\ }\textbf {\bibinfo {volume} {47}},\ \bibinfo {pages} {558--561} (\bibinfo {year} {1993})}\BibitemShut {NoStop}%
\bibitem [{\citenamefont {Kresse}\ and\ \citenamefont {Joubert}(1999)}]{Kresse1999PAW}%
  \BibitemOpen
  \bibfield  {author} {\bibinfo {author} {\bibfnamefont {G.}~\bibnamefont {Kresse}}\ and\ \bibinfo {author} {\bibfnamefont {D.}~\bibnamefont {Joubert}},\ }\bibfield  {title} {\enquote {\bibinfo {title} {From ultrasoft pseudopotentials to the projector augmented-wave method},}\ }\href {https://doi.org/10.1103/PhysRevB.59.1758} {\bibfield  {journal} {\bibinfo  {journal} {Phys. Rev. B}\ }\textbf {\bibinfo {volume} {59}},\ \bibinfo {pages} {1758--1775} (\bibinfo {year} {1999})}\BibitemShut {NoStop}%
\bibitem [{\citenamefont {Perdew}\ \emph {et~al.}(2008)\citenamefont {Perdew}, \citenamefont {Ruzsinszky}, \citenamefont {Csonka}, \citenamefont {Vydrov}, \citenamefont {Scuseria}, \citenamefont {Constantin}, \citenamefont {Zhou},\ and\ \citenamefont {Burke}}]{Perdew2008PBE}%
  \BibitemOpen
  \bibfield  {author} {\bibinfo {author} {\bibfnamefont {J.~P.}\ \bibnamefont {Perdew}}, \bibinfo {author} {\bibfnamefont {A.}~\bibnamefont {Ruzsinszky}}, \bibinfo {author} {\bibfnamefont {G.~I.}\ \bibnamefont {Csonka}}, \bibinfo {author} {\bibfnamefont {O.~A.}\ \bibnamefont {Vydrov}}, \bibinfo {author} {\bibfnamefont {G.~E.}\ \bibnamefont {Scuseria}}, \bibinfo {author} {\bibfnamefont {L.~A.}\ \bibnamefont {Constantin}}, \bibinfo {author} {\bibfnamefont {X.}~\bibnamefont {Zhou}},\ and\ \bibinfo {author} {\bibfnamefont {K.}~\bibnamefont {Burke}},\ }\bibfield  {title} {\enquote {\bibinfo {title} {Restoring the density-gradient expansion for exchange in solids and surfaces},}\ }\href {https://doi.org/10.1103/PhysRevLett.100.136406} {\bibfield  {journal} {\bibinfo  {journal} {Phys. Rev. Lett.}\ }\textbf {\bibinfo {volume} {100}},\ \bibinfo {pages} {136406} (\bibinfo {year} {2008})}\BibitemShut {NoStop}%
\bibitem [{\citenamefont {Ohnoutek}\ \emph {et~al.}(2016)\citenamefont {Ohnoutek}, \citenamefont {Hakl}, \citenamefont {Veis}, \citenamefont {Piot}, \citenamefont {Faugeras}, \citenamefont {Martinez}, \citenamefont {Yakushev}, \citenamefont {Martin}, \citenamefont {Dra{\v{s}}ar}, \citenamefont {Materna} \emph {et~al.}}]{strongFaradayScirep}%
  \BibitemOpen
  \bibfield  {author} {\bibinfo {author} {\bibfnamefont {L.}~\bibnamefont {Ohnoutek}}, \bibinfo {author} {\bibfnamefont {M.}~\bibnamefont {Hakl}}, \bibinfo {author} {\bibfnamefont {M.}~\bibnamefont {Veis}}, \bibinfo {author} {\bibfnamefont {B.}~\bibnamefont {Piot}}, \bibinfo {author} {\bibfnamefont {C.}~\bibnamefont {Faugeras}}, \bibinfo {author} {\bibfnamefont {G.}~\bibnamefont {Martinez}}, \bibinfo {author} {\bibfnamefont {M.}~\bibnamefont {Yakushev}}, \bibinfo {author} {\bibfnamefont {R.}~\bibnamefont {Martin}}, \bibinfo {author} {\bibfnamefont {{\v{C}}.}~\bibnamefont {Dra{\v{s}}ar}}, \bibinfo {author} {\bibfnamefont {A.}~\bibnamefont {Materna}}, \emph {et~al.},\ }\bibfield  {title} {\enquote {\bibinfo {title} {Strong interband {F}araday rotation in 3d topological insulator $\mathrm{Bi_2Se_3}$},}\ }\href@noop {} {\bibfield  {journal} {\bibinfo  {journal} {Sci. Rep.}\ }\textbf {\bibinfo {volume} {6}},\ \bibinfo {pages} {19087} (\bibinfo {year} {2016})}\BibitemShut {NoStop}%
\bibitem [{\citenamefont {Zhang}\ \emph {et~al.}(2010)\citenamefont {Zhang}, \citenamefont {He}, \citenamefont {Chang}, \citenamefont {Song}, \citenamefont {Wang}, \citenamefont {Chen}, \citenamefont {Jia}, \citenamefont {Fang}, \citenamefont {Dai}, \citenamefont {Shan} \emph {et~al.}}]{zhang2010crossover}%
  \BibitemOpen
  \bibfield  {author} {\bibinfo {author} {\bibfnamefont {Y.}~\bibnamefont {Zhang}}, \bibinfo {author} {\bibfnamefont {K.}~\bibnamefont {He}}, \bibinfo {author} {\bibfnamefont {C.-Z.}\ \bibnamefont {Chang}}, \bibinfo {author} {\bibfnamefont {C.-L.}\ \bibnamefont {Song}}, \bibinfo {author} {\bibfnamefont {L.-L.}\ \bibnamefont {Wang}}, \bibinfo {author} {\bibfnamefont {X.}~\bibnamefont {Chen}}, \bibinfo {author} {\bibfnamefont {J.-F.}\ \bibnamefont {Jia}}, \bibinfo {author} {\bibfnamefont {Z.}~\bibnamefont {Fang}}, \bibinfo {author} {\bibfnamefont {X.}~\bibnamefont {Dai}}, \bibinfo {author} {\bibfnamefont {W.-Y.}\ \bibnamefont {Shan}}, \emph {et~al.},\ }\bibfield  {title} {\enquote {\bibinfo {title} {Crossover of the three-dimensional topological insulator $\mathrm{Bi_2Se_3}$ to the two-dimensional limit},}\ }\href@noop {} {\bibfield  {journal} {\bibinfo  {journal} {Nat. Phys.}\ }\textbf {\bibinfo {volume} {6}},\ \bibinfo {pages} {584} (\bibinfo {year} {2010})}\BibitemShut {NoStop}%
\bibitem [{\citenamefont {Edmonds}\ \emph {et~al.}(2014)\citenamefont {Edmonds}, \citenamefont {Hellerstedt}, \citenamefont {Tadich}, \citenamefont {Schenk}, \citenamefont {O’Donnell}, \citenamefont {Tosado}, \citenamefont {Butch}, \citenamefont {Syers}, \citenamefont {Paglione},\ and\ \citenamefont {Fuhrer}}]{edmonds2014stability}%
  \BibitemOpen
  \bibfield  {author} {\bibinfo {author} {\bibfnamefont {M.~T.}\ \bibnamefont {Edmonds}}, \bibinfo {author} {\bibfnamefont {J.~T.}\ \bibnamefont {Hellerstedt}}, \bibinfo {author} {\bibfnamefont {A.}~\bibnamefont {Tadich}}, \bibinfo {author} {\bibfnamefont {A.}~\bibnamefont {Schenk}}, \bibinfo {author} {\bibfnamefont {K.~M.}\ \bibnamefont {O’Donnell}}, \bibinfo {author} {\bibfnamefont {J.}~\bibnamefont {Tosado}}, \bibinfo {author} {\bibfnamefont {N.~P.}\ \bibnamefont {Butch}}, \bibinfo {author} {\bibfnamefont {P.}~\bibnamefont {Syers}}, \bibinfo {author} {\bibfnamefont {J.}~\bibnamefont {Paglione}},\ and\ \bibinfo {author} {\bibfnamefont {M.~S.}\ \bibnamefont {Fuhrer}},\ }\bibfield  {title} {\enquote {\bibinfo {title} {Stability and surface reconstruction of topological insulator $\mathrm{Bi_2Se_3}$ on exposure to atmosphere},}\ }\href@noop {} {\bibfield  {journal} {\bibinfo  {journal} {J. Phys. Chem. C}\ }\textbf {\bibinfo {volume} {118}},\ \bibinfo {pages} {20413} (\bibinfo {year} {2014})}\BibitemShut
  {NoStop}%
\bibitem [{\citenamefont {Reim}\ and\ \citenamefont {Schoenes}(1990)}]{FerromagmattHandbook}%
  \BibitemOpen
  \bibfield  {author} {\bibinfo {author} {\bibfnamefont {W.}~\bibnamefont {Reim}}\ and\ \bibinfo {author} {\bibfnamefont {J.}~\bibnamefont {Schoenes}},\ }\href@noop {} {\emph {\bibinfo {title} {{Ferromagnetic Materials. A Handbook on the Properties of Magnetically Ordered Substances}}}},\ edited by\ \bibinfo {editor} {\bibfnamefont {K.~H.~J.}\ \bibnamefont {Buschow}}\ and\ \bibinfo {editor} {\bibfnamefont {E.~P.}\ \bibnamefont {Wohlfahrt}},\ Vol.~\bibinfo {volume} {5}\ (\bibinfo  {publisher} {North--Holland},\ \bibinfo {year} {1990})\ Chap.~\bibinfo {chapter} {2}\BibitemShut {NoStop}%
\bibitem [{\citenamefont {Shinagawa}(1999)}]{MagnetoOptics}%
  \BibitemOpen
  \bibfield  {author} {\bibinfo {author} {\bibfnamefont {K.}~\bibnamefont {Shinagawa}},\ }\href@noop {} {\emph {\bibinfo {title} {Magneto-{O}ptics}}},\ edited by\ \bibinfo {editor} {\bibfnamefont {S.}~\bibnamefont {Sugano}}\ and\ \bibinfo {editor} {\bibfnamefont {N.}~\bibnamefont {Kojima}},\ {SOLID-STATE SCIENCES}\ (\bibinfo  {publisher} {Springer},\ \bibinfo {year} {1999})\ Chap.~\bibinfo {chapter} {5}\BibitemShut {NoStop}%
\bibitem [{\citenamefont {Zhang}\ \emph {et~al.}(2011)\citenamefont {Zhang}, \citenamefont {Peng}, \citenamefont {Soni}, \citenamefont {Zhao}, \citenamefont {Xiong}, \citenamefont {Peng}, \citenamefont {Wang}, \citenamefont {Dresselhaus},\ and\ \citenamefont {Xiong}}]{BiSeRamanNanolett2011}%
  \BibitemOpen
  \bibfield  {author} {\bibinfo {author} {\bibfnamefont {J.}~\bibnamefont {Zhang}}, \bibinfo {author} {\bibfnamefont {Z.}~\bibnamefont {Peng}}, \bibinfo {author} {\bibfnamefont {A.}~\bibnamefont {Soni}}, \bibinfo {author} {\bibfnamefont {Y.}~\bibnamefont {Zhao}}, \bibinfo {author} {\bibfnamefont {Y.}~\bibnamefont {Xiong}}, \bibinfo {author} {\bibfnamefont {B.}~\bibnamefont {Peng}}, \bibinfo {author} {\bibfnamefont {J.}~\bibnamefont {Wang}}, \bibinfo {author} {\bibfnamefont {M.~S.}\ \bibnamefont {Dresselhaus}},\ and\ \bibinfo {author} {\bibfnamefont {Q.}~\bibnamefont {Xiong}},\ }\bibfield  {title} {\enquote {\bibinfo {title} {Raman {S}pectroscopy of {F}ew-{Q}uintuple {L}ayer {T}opological {I}nsulator $\mathrm{Bi_2Se_3}$ {N}anoplatelets},}\ }\href@noop {} {\bibfield  {journal} {\bibinfo  {journal} {Nano Lett.}\ }\textbf {\bibinfo {volume} {11}},\ \bibinfo {pages} {2407} (\bibinfo {year} {2011})}\BibitemShut {NoStop}%
\bibitem [{\citenamefont {Norimatsu}\ \emph {et~al.}(2013)\citenamefont {Norimatsu}, \citenamefont {Hu}, \citenamefont {Goto}, \citenamefont {Igarashi}, \citenamefont {Sasagawa},\ and\ \citenamefont {Nakamura}}]{norimatsu2013SSC}%
  \BibitemOpen
  \bibfield  {author} {\bibinfo {author} {\bibfnamefont {K.}~\bibnamefont {Norimatsu}}, \bibinfo {author} {\bibfnamefont {J.}~\bibnamefont {Hu}}, \bibinfo {author} {\bibfnamefont {A.}~\bibnamefont {Goto}}, \bibinfo {author} {\bibfnamefont {K.}~\bibnamefont {Igarashi}}, \bibinfo {author} {\bibfnamefont {T.}~\bibnamefont {Sasagawa}},\ and\ \bibinfo {author} {\bibfnamefont {K.~G.}\ \bibnamefont {Nakamura}},\ }\bibfield  {title} {\enquote {\bibinfo {title} {Coherent optical phonons in a $\mathrm{Bi_2Se_3}$ single crystal measured via transient anisotropic reflectivity},}\ }\href@noop {} {\bibfield  {journal} {\bibinfo  {journal} {Solid State Commun.}\ }\textbf {\bibinfo {volume} {157}},\ \bibinfo {pages} {58} (\bibinfo {year} {2013})}\BibitemShut {NoStop}%
\bibitem [{\citenamefont {Norimatsu}\ \emph {et~al.}(2015)\citenamefont {Norimatsu}, \citenamefont {Hada}, \citenamefont {Yamamoto}, \citenamefont {Sasagawa}, \citenamefont {Kitajima}, \citenamefont {Kayanuma},\ and\ \citenamefont {Nakamura}}]{norimatsu2015allphonon}%
  \BibitemOpen
  \bibfield  {author} {\bibinfo {author} {\bibfnamefont {K.}~\bibnamefont {Norimatsu}}, \bibinfo {author} {\bibfnamefont {M.}~\bibnamefont {Hada}}, \bibinfo {author} {\bibfnamefont {S.}~\bibnamefont {Yamamoto}}, \bibinfo {author} {\bibfnamefont {T.}~\bibnamefont {Sasagawa}}, \bibinfo {author} {\bibfnamefont {M.}~\bibnamefont {Kitajima}}, \bibinfo {author} {\bibfnamefont {Y.}~\bibnamefont {Kayanuma}},\ and\ \bibinfo {author} {\bibfnamefont {K.~G.}\ \bibnamefont {Nakamura}},\ }\bibfield  {title} {\enquote {\bibinfo {title} {Dynamics of all the {R}aman-active coherent phonons in $\mathrm{Sb_2Te_3}$ revealed via transient reflectivity},}\ }\href@noop {} {\bibfield  {journal} {\bibinfo  {journal} {J. Appl. Phys.}\ }\textbf {\bibinfo {volume} {117}} (\bibinfo {year} {2015})}\BibitemShut {NoStop}%
\bibitem [{\citenamefont {Sie}\ \emph {et~al.}(2015)\citenamefont {Sie}, \citenamefont {McIver}, \citenamefont {Lee}, \citenamefont {Fu}, \citenamefont {Kong},\ and\ \citenamefont {Gedik}}]{OSENatMat2015}%
  \BibitemOpen
  \bibfield  {author} {\bibinfo {author} {\bibfnamefont {E.~J.}\ \bibnamefont {Sie}}, \bibinfo {author} {\bibfnamefont {J.~W.}\ \bibnamefont {McIver}}, \bibinfo {author} {\bibfnamefont {Y.~H.}\ \bibnamefont {Lee}}, \bibinfo {author} {\bibfnamefont {L.}~\bibnamefont {Fu}}, \bibinfo {author} {\bibfnamefont {J.}~\bibnamefont {Kong}},\ and\ \bibinfo {author} {\bibfnamefont {N.}~\bibnamefont {Gedik}},\ }\bibfield  {title} {\enquote {\bibinfo {title} {Valley-selective optical {S}tark effect in monolayer $\mathrm{WS_2}$},}\ }\href@noop {} {\bibfield  {journal} {\bibinfo  {journal} {Nat. Mater.}\ }\textbf {\bibinfo {volume} {14}},\ \bibinfo {pages} {290} (\bibinfo {year} {2015})}\BibitemShut {NoStop}%
\bibitem [{\citenamefont {LaMountain}\ \emph {et~al.}(2018)\citenamefont {LaMountain}, \citenamefont {Bergeron}, \citenamefont {Balla}, \citenamefont {Stanev}, \citenamefont {Hersam},\ and\ \citenamefont {Stern}}]{OSEprobedKerr_PRB}%
  \BibitemOpen
  \bibfield  {author} {\bibinfo {author} {\bibfnamefont {T.}~\bibnamefont {LaMountain}}, \bibinfo {author} {\bibfnamefont {H.}~\bibnamefont {Bergeron}}, \bibinfo {author} {\bibfnamefont {I.}~\bibnamefont {Balla}}, \bibinfo {author} {\bibfnamefont {T.~K.}\ \bibnamefont {Stanev}}, \bibinfo {author} {\bibfnamefont {M.~C.}\ \bibnamefont {Hersam}},\ and\ \bibinfo {author} {\bibfnamefont {N.~P.}\ \bibnamefont {Stern}},\ }\bibfield  {title} {\enquote {\bibinfo {title} {Valley-selective optical {S}tark effect probed by {K}err rotation},}\ }\href {https://doi.org/10.1103/PhysRevB.97.045307} {\bibfield  {journal} {\bibinfo  {journal} {Phys. Rev. B}\ }\textbf {\bibinfo {volume} {97}},\ \bibinfo {pages} {045307} (\bibinfo {year} {2018})}\BibitemShut {NoStop}%
\bibitem [{\citenamefont {Chernikov}\ \emph {et~al.}(2015{\natexlab{b}})\citenamefont {Chernikov}, \citenamefont {Ruppert}, \citenamefont {Hill}, \citenamefont {Rigosi},\ and\ \citenamefont {Heinz}}]{BGRNatPhoto2015}%
  \BibitemOpen
  \bibfield  {author} {\bibinfo {author} {\bibfnamefont {A.}~\bibnamefont {Chernikov}}, \bibinfo {author} {\bibfnamefont {C.}~\bibnamefont {Ruppert}}, \bibinfo {author} {\bibfnamefont {H.~M.}\ \bibnamefont {Hill}}, \bibinfo {author} {\bibfnamefont {A.~F.}\ \bibnamefont {Rigosi}},\ and\ \bibinfo {author} {\bibfnamefont {T.~F.}\ \bibnamefont {Heinz}},\ }\bibfield  {title} {\enquote {\bibinfo {title} {Population inversion and giant bandgap renormalization in atomically thin $\mathrm{WS_2}$ layers},}\ }\href@noop {} {\bibfield  {journal} {\bibinfo  {journal} {Nat. Photonics}\ }\textbf {\bibinfo {volume} {9}},\ \bibinfo {pages} {466} (\bibinfo {year} {2015}{\natexlab{b}})}\BibitemShut {NoStop}%
\bibitem [{\citenamefont {Henn}\ \emph {et~al.}(2013)\citenamefont {Henn}, \citenamefont {Heckel}, \citenamefont {Beck}, \citenamefont {Kiessling}, \citenamefont {Ossau}, \citenamefont {Molenkamp}, \citenamefont {Reuter},\ and\ \citenamefont {Wieck}}]{HorKerrPRB2013}%
  \BibitemOpen
  \bibfield  {author} {\bibinfo {author} {\bibfnamefont {T.}~\bibnamefont {Henn}}, \bibinfo {author} {\bibfnamefont {A.}~\bibnamefont {Heckel}}, \bibinfo {author} {\bibfnamefont {M.}~\bibnamefont {Beck}}, \bibinfo {author} {\bibfnamefont {T.}~\bibnamefont {Kiessling}}, \bibinfo {author} {\bibfnamefont {W.}~\bibnamefont {Ossau}}, \bibinfo {author} {\bibfnamefont {L.~W.}\ \bibnamefont {Molenkamp}}, \bibinfo {author} {\bibfnamefont {D.}~\bibnamefont {Reuter}},\ and\ \bibinfo {author} {\bibfnamefont {A.~D.}\ \bibnamefont {Wieck}},\ }\bibfield  {title} {\enquote {\bibinfo {title} {Hot carrier effects on the magneto-optical detection of electron spins in gaas},}\ }\href {https://doi.org/10.1103/PhysRevB.88.085303} {\bibfield  {journal} {\bibinfo  {journal} {Phys. Rev. B}\ }\textbf {\bibinfo {volume} {88}},\ \bibinfo {pages} {085303} (\bibinfo {year} {2013})}\BibitemShut {NoStop}%
\bibitem [{\citenamefont {Shang}\ \emph {et~al.}(2020)\citenamefont {Shang}, \citenamefont {Feng}, \citenamefont {Zhao}, \citenamefont {Li}, \citenamefont {Pan},\ and\ \citenamefont {Zhao}}]{shang2020saturable}%
  \BibitemOpen
  \bibfield  {author} {\bibinfo {author} {\bibfnamefont {J.}~\bibnamefont {Shang}}, \bibinfo {author} {\bibfnamefont {T.}~\bibnamefont {Feng}}, \bibinfo {author} {\bibfnamefont {S.}~\bibnamefont {Zhao}}, \bibinfo {author} {\bibfnamefont {T.}~\bibnamefont {Li}}, \bibinfo {author} {\bibfnamefont {Z.}~\bibnamefont {Pan}},\ and\ \bibinfo {author} {\bibfnamefont {J.}~\bibnamefont {Zhao}},\ }\bibfield  {title} {\enquote {\bibinfo {title} {Saturable absorption characteristics of $\mathrm{Bi_2Se_3}$ in a 2 $\mu$m {Q}-switching bulk laser},}\ }\href@noop {} {\bibfield  {journal} {\bibinfo  {journal} {Opt. Express}\ }\textbf {\bibinfo {volume} {28}},\ \bibinfo {pages} {5639} (\bibinfo {year} {2020})}\BibitemShut {NoStop}%
\bibitem [{\citenamefont {Mori}\ \emph {et~al.}(2023)\citenamefont {Mori}, \citenamefont {Ciocys}, \citenamefont {Takasan}, \citenamefont {Ai}, \citenamefont {Currier}, \citenamefont {Morimoto}, \citenamefont {Moore},\ and\ \citenamefont {Lanzara}}]{mori2023spatiallyNature}%
  \BibitemOpen
  \bibfield  {author} {\bibinfo {author} {\bibfnamefont {R.}~\bibnamefont {Mori}}, \bibinfo {author} {\bibfnamefont {S.}~\bibnamefont {Ciocys}}, \bibinfo {author} {\bibfnamefont {K.}~\bibnamefont {Takasan}}, \bibinfo {author} {\bibfnamefont {P.}~\bibnamefont {Ai}}, \bibinfo {author} {\bibfnamefont {K.}~\bibnamefont {Currier}}, \bibinfo {author} {\bibfnamefont {T.}~\bibnamefont {Morimoto}}, \bibinfo {author} {\bibfnamefont {J.~E.}\ \bibnamefont {Moore}},\ and\ \bibinfo {author} {\bibfnamefont {A.}~\bibnamefont {Lanzara}},\ }\bibfield  {title} {\enquote {\bibinfo {title} {Spin-polarized spatially indirect excitons in a topological insulator},}\ }\href@noop {} {\bibfield  {journal} {\bibinfo  {journal} {Nature}\ }\textbf {\bibinfo {volume} {614}},\ \bibinfo {pages} {249} (\bibinfo {year} {2023})}\BibitemShut {NoStop}%
\bibitem [{\citenamefont {Jiang}\ \emph {et~al.}(2017)\citenamefont {Jiang}, \citenamefont {Liu}, \citenamefont {Li},\ and\ \citenamefont {Duan}}]{PRLScaling2017}%
  \BibitemOpen
  \bibfield  {author} {\bibinfo {author} {\bibfnamefont {Z.}~\bibnamefont {Jiang}}, \bibinfo {author} {\bibfnamefont {Z.}~\bibnamefont {Liu}}, \bibinfo {author} {\bibfnamefont {Y.}~\bibnamefont {Li}},\ and\ \bibinfo {author} {\bibfnamefont {W.}~\bibnamefont {Duan}},\ }\bibfield  {title} {\enquote {\bibinfo {title} {Scaling universality between band gap and exciton binding energy of two-dimensional semiconductors},}\ }\href {https://doi.org/10.1103/PhysRevLett.118.266401} {\bibfield  {journal} {\bibinfo  {journal} {Phys. Rev. Lett.}\ }\textbf {\bibinfo {volume} {118}},\ \bibinfo {pages} {266401} (\bibinfo {year} {2017})}\BibitemShut {NoStop}%
\bibitem [{\citenamefont {Ou}\ \emph {et~al.}(2014)\citenamefont {Ou}, \citenamefont {So}, \citenamefont {Adamo}, \citenamefont {Sulaev}, \citenamefont {Wang},\ and\ \citenamefont {Zheludev}}]{Ou2014ultravioletNatCom}%
  \BibitemOpen
  \bibfield  {author} {\bibinfo {author} {\bibfnamefont {J.-Y.}\ \bibnamefont {Ou}}, \bibinfo {author} {\bibfnamefont {J.-K.}\ \bibnamefont {So}}, \bibinfo {author} {\bibfnamefont {G.}~\bibnamefont {Adamo}}, \bibinfo {author} {\bibfnamefont {A.}~\bibnamefont {Sulaev}}, \bibinfo {author} {\bibfnamefont {L.}~\bibnamefont {Wang}},\ and\ \bibinfo {author} {\bibfnamefont {N.~I.}\ \bibnamefont {Zheludev}},\ }\bibfield  {title} {\enquote {\bibinfo {title} {Ultraviolet and visible range plasmonics in the topological insulator $\mathrm{Bi_{1.5}Sb_{0. 5}Te_{1.8}Se_{1.2}}$},}\ }\href@noop {} {\bibfield  {journal} {\bibinfo  {journal} {Nat. Commun.}\ }\textbf {\bibinfo {volume} {5}},\ \bibinfo {pages} {5139} (\bibinfo {year} {2014})}\BibitemShut {NoStop}%
\end{thebibliography}%


\providecommand{\noopsort}[1]{}\providecommand{\singleletter}[1]{#1}%
\begin{thebibliography}{3}%
\makeatletter
\providecommand \@ifxundefined [1]{%
 \@ifx{#1\undefined}
}%
\providecommand \@ifnum [1]{%
 \ifnum #1\expandafter \@firstoftwo
 \else \expandafter \@secondoftwo
 \fi
}%
\providecommand \@ifx [1]{%
 \ifx #1\expandafter \@firstoftwo
 \else \expandafter \@secondoftwo
 \fi
}%
\providecommand \natexlab [1]{#1}%
\providecommand \enquote  [1]{``#1''}%
\providecommand \bibnamefont  [1]{#1}%
\providecommand \bibfnamefont [1]{#1}%
\providecommand \citenamefont [1]{#1}%
\providecommand \href@noop [0]{\@secondoftwo}%
\providecommand \href [0]{\begingroup \@sanitize@url \@href}%
\providecommand \@href[1]{\@@startlink{#1}\@@href}%
\providecommand \@@href[1]{\endgroup#1\@@endlink}%
\providecommand \@sanitize@url [0]{\catcode `\\12\catcode `\$12\catcode `\&12\catcode `\#12\catcode `\^12\catcode `\_12\catcode `\%12\relax}%
\providecommand \@@startlink[1]{}%
\providecommand \@@endlink[0]{}%
\providecommand \url  [0]{\begingroup\@sanitize@url \@url }%
\providecommand \@url [1]{\endgroup\@href {#1}{\urlprefix }}%
\providecommand \urlprefix  [0]{URL }%
\providecommand \Eprint [0]{\href }%
\providecommand \doibase [0]{https://doi.org/}%
\providecommand \selectlanguage [0]{\@gobble}%
\providecommand \bibinfo  [0]{\@secondoftwo}%
\providecommand \bibfield  [0]{\@secondoftwo}%
\providecommand \translation [1]{[#1]}%
\providecommand \BibitemOpen [0]{}%
\providecommand \bibitemStop [0]{}%
\providecommand \bibitemNoStop [0]{.\EOS\space}%
\providecommand \EOS [0]{\spacefactor3000\relax}%
\providecommand \BibitemShut  [1]{\csname bibitem#1\endcsname}%
\let\auto@bib@innerbib\@empty
\bibitem [{\citenamefont {Richter}\ and\ \citenamefont {Becker}(1977)}]{richter1977raman}%
  \BibitemOpen
  \bibfield  {author} {\bibinfo {author} {\bibfnamefont {W.}~\bibnamefont {Richter}}\ and\ \bibinfo {author} {\bibfnamefont {C.}~\bibnamefont {Becker}},\ }\bibfield  {title} {\enquote {\bibinfo {title} {A {R}aman and far-infrared investigation of phonons in the rhombohedral $\mathrm{V_{2}-VI_{3}}$ compounds $\mathrm{Bi_{2}Te_{3}}$, $\mathrm{Bi_{2}Se_{3}}$, $\mathrm{Sb_{2}Te_{3}}$ and $\mathrm{Bi}_{2} (\mathrm{Te}_{1-x}\mathrm{Se}_x)_3\ (0<x<1),\ (\mathrm{Bi}_{1-y}\mathrm{Sb}_y)_2\mathrm{Te}_3\ (0< y< 1)$},}\ }\href@noop {} {\bibfield  {journal} {\bibinfo  {journal} {Phys. Status Solidi (b)}\ }\textbf {\bibinfo {volume} {84}},\ \bibinfo {pages} {619--628} (\bibinfo {year} {1977})}\BibitemShut {NoStop}%
\bibitem [{\citenamefont {Zhang}\ \emph {et~al.}(2011)\citenamefont {Zhang}, \citenamefont {Peng}, \citenamefont {Soni}, \citenamefont {Zhao}, \citenamefont {Xiong}, \citenamefont {Peng}, \citenamefont {Wang}, \citenamefont {Dresselhaus},\ and\ \citenamefont {Xiong}}]{BiSeRamanNanolett2011}%
  \BibitemOpen
  \bibfield  {author} {\bibinfo {author} {\bibfnamefont {J.}~\bibnamefont {Zhang}}, \bibinfo {author} {\bibfnamefont {Z.}~\bibnamefont {Peng}}, \bibinfo {author} {\bibfnamefont {A.}~\bibnamefont {Soni}}, \bibinfo {author} {\bibfnamefont {Y.}~\bibnamefont {Zhao}}, \bibinfo {author} {\bibfnamefont {Y.}~\bibnamefont {Xiong}}, \bibinfo {author} {\bibfnamefont {B.}~\bibnamefont {Peng}}, \bibinfo {author} {\bibfnamefont {J.}~\bibnamefont {Wang}}, \bibinfo {author} {\bibfnamefont {M.~S.}\ \bibnamefont {Dresselhaus}},\ and\ \bibinfo {author} {\bibfnamefont {Q.}~\bibnamefont {Xiong}},\ }\bibfield  {title} {\enquote {\bibinfo {title} {Raman {S}pectroscopy of {F}ew-{Q}uintuple {L}ayer {T}opological {I}nsulator $\mathrm{Bi_2Se_3}$ {N}anoplatelets},}\ }\href@noop {} {\bibfield  {journal} {\bibinfo  {journal} {Nano Lett.}\ }\textbf {\bibinfo {volume} {11}},\ \bibinfo {pages} {2407} (\bibinfo {year} {2011})}\BibitemShut {NoStop}%
\bibitem [{\citenamefont {Norimatsu}\ \emph {et~al.}(2015)\citenamefont {Norimatsu}, \citenamefont {Hada}, \citenamefont {Yamamoto}, \citenamefont {Sasagawa}, \citenamefont {Kitajima}, \citenamefont {Kayanuma},\ and\ \citenamefont {Nakamura}}]{norimatsu2015allphonon}%
  \BibitemOpen
  \bibfield  {author} {\bibinfo {author} {\bibfnamefont {K.}~\bibnamefont {Norimatsu}}, \bibinfo {author} {\bibfnamefont {M.}~\bibnamefont {Hada}}, \bibinfo {author} {\bibfnamefont {S.}~\bibnamefont {Yamamoto}}, \bibinfo {author} {\bibfnamefont {T.}~\bibnamefont {Sasagawa}}, \bibinfo {author} {\bibfnamefont {M.}~\bibnamefont {Kitajima}}, \bibinfo {author} {\bibfnamefont {Y.}~\bibnamefont {Kayanuma}},\ and\ \bibinfo {author} {\bibfnamefont {K.~G.}\ \bibnamefont {Nakamura}},\ }\bibfield  {title} {\enquote {\bibinfo {title} {Dynamics of all the {R}aman-active coherent phonons in $\mathrm{Sb_2Te_3}$ revealed via transient reflectivity},}\ }\href@noop {} {\bibfield  {journal} {\bibinfo  {journal} {J. Appl. Phys.}\ }\textbf {\bibinfo {volume} {117}} (\bibinfo {year} {2015})}\BibitemShut {NoStop}%
\end{thebibliography}%

\begin{figure}[p]
    \centering
    \includegraphics[width = 17cm]{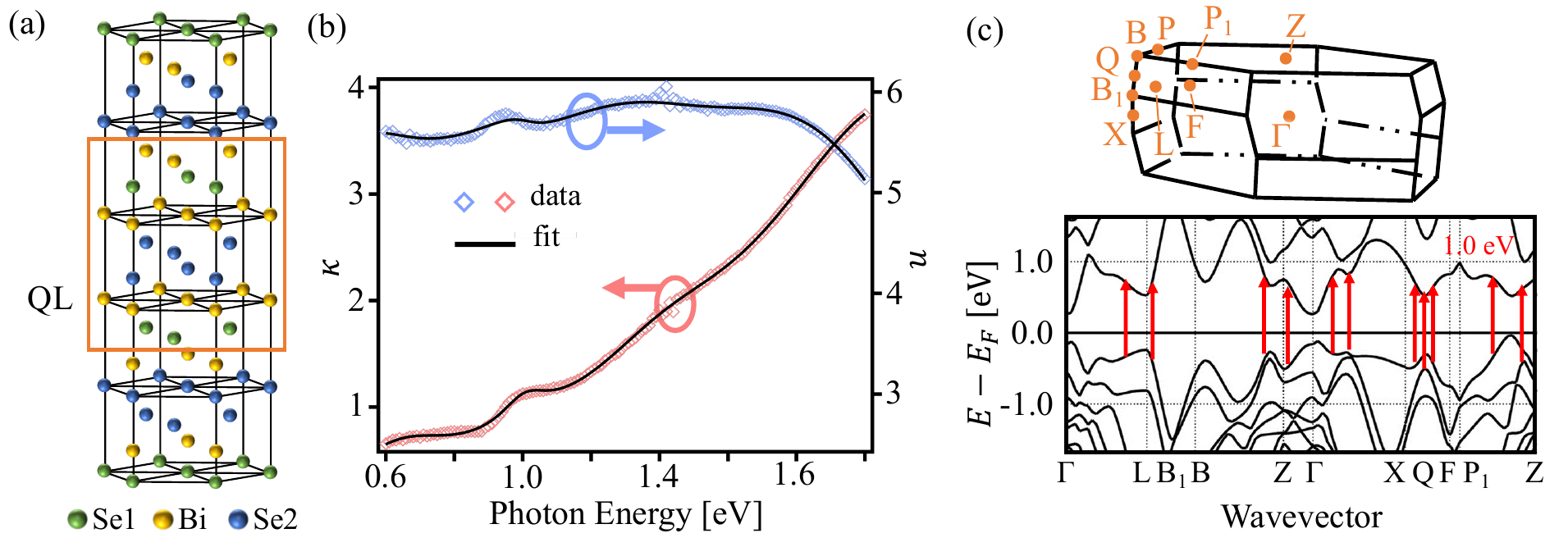}
    \caption{
    (a) Schematic of the crystal structure of $\mathrm{Bi_{2}Se_{3}}$. QL represents the quintuple layer. (b) The measured imaginary part($\kappa$) and real($n$) part of the complex dielectric function($\hat{n}$) of $\mathrm{Bi_{2}Se_{3}}$. (c) (upper side) The Brillouin zone of $\mathrm{Bi_2Se_3}$. (lower side) Calculated band structure of the bulk $\mathrm{Bi_{2}Se_{3}}$. The red arrows indicate possible optical transitions by 1.0 eV light. These transitions occurred at $\Gamma$ and $Z$ point.
    }
    \label{figure1}
\end{figure}

\begin{figure}[p]
    \centering
    \includegraphics[width = 17cm]{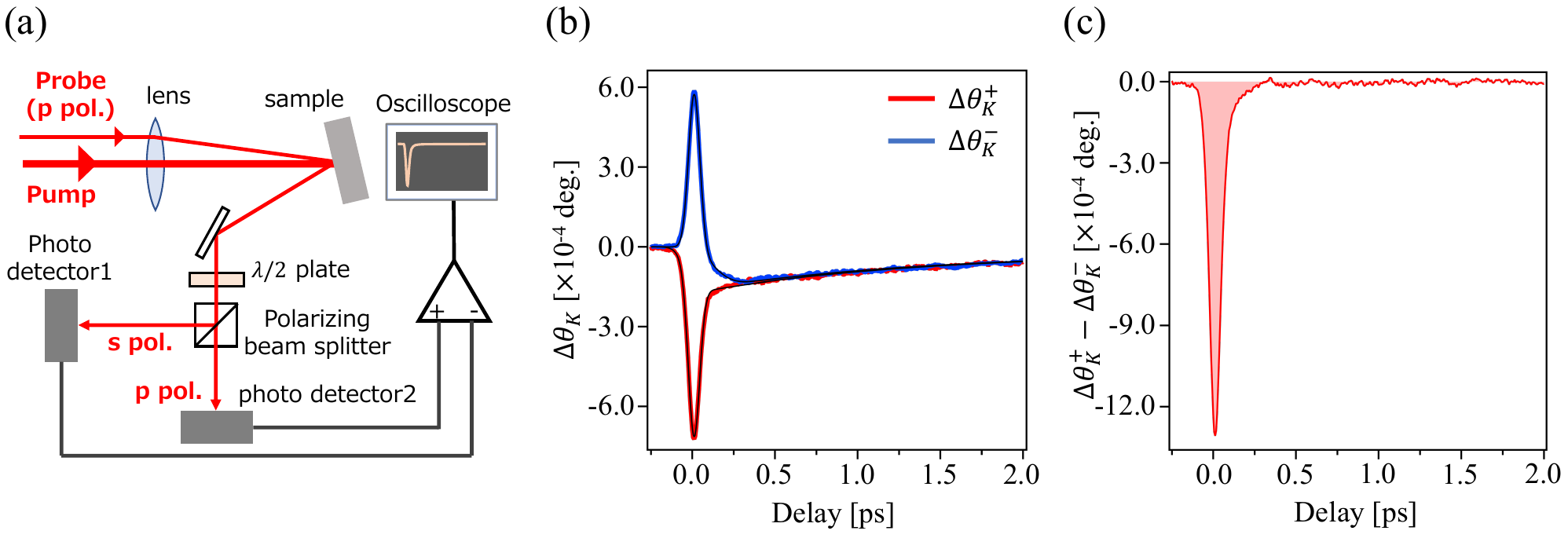}
    \caption{
    (a) Schematic of the experimental setup for time resolved-Kerr rotation measurement. (b) A time trace of transient Kerr rotation ($\Delta\theta_{K}^{\pm}$) induced by right and left circularly polarized pump at 0.95 eV. The sign of transient zero-time-delay peak reverses depending on the helicity. (c) A time trace of subtracted signal $[\Delta\theta_{K}^{+}-\Delta\theta_{K}^{-}]$ at 0.95 eV.
    }

    \label{figure2}
\end{figure}

\begin{figure}[p]
    \centering
    \includegraphics[width = 8.5cm]{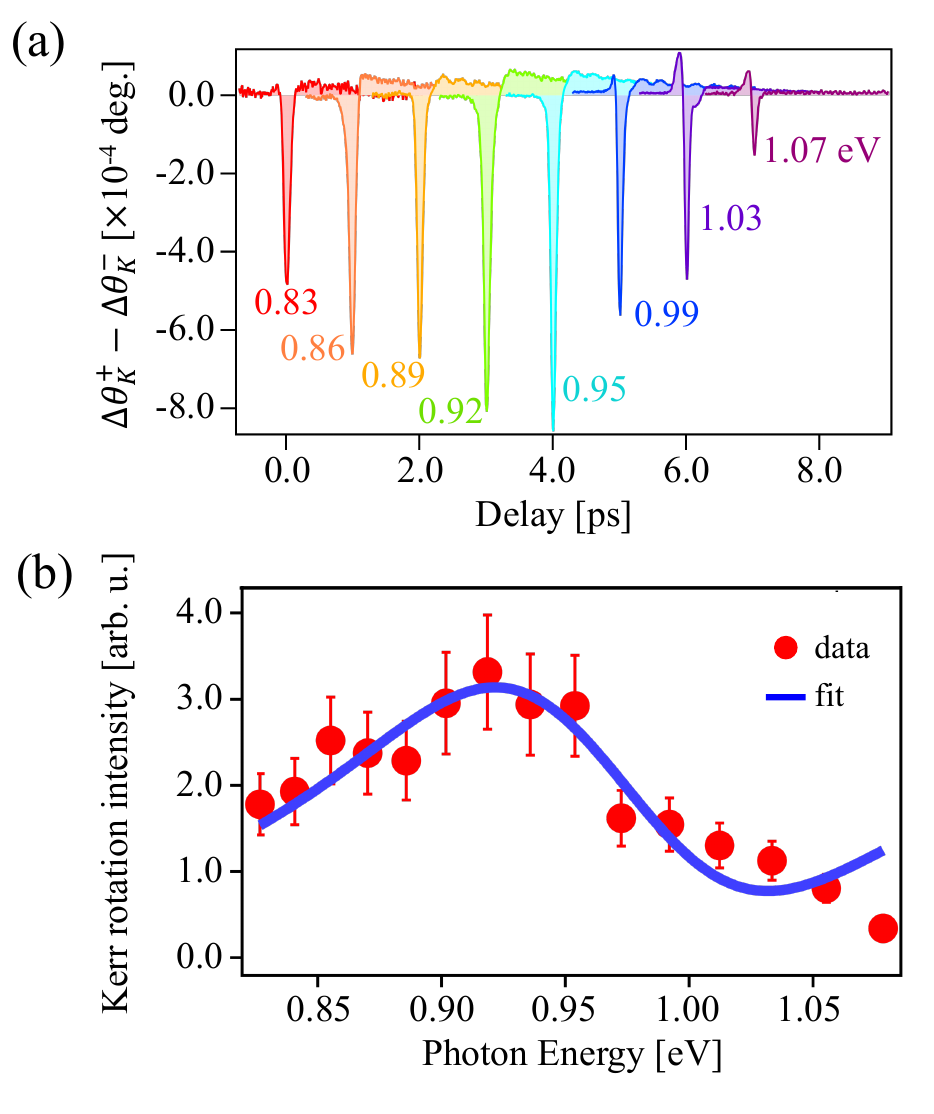}
    \caption{
    (a) Photon-energy dependence of $\Delta\theta_{K}^{+}-\Delta\theta_{K}^{-}$ observed at constant incident pump fluence 0.5 $\mathrm{mJ/cm^{2}}$. (b) Photon-energy dependence of area of zero-time delay peak (Kerr rotation intensity). The error bars are set to ±20 \% of intensity. The data are fit by equation (\ref{Kerrrotation_kappa}). 
    }
    \label{figure3}
\end{figure}

\end{document}


\title{
Supplementary Materials for\\
"Spin dependent bandgap renormalization and state filling effect in Bi$_2$Se$_3$ observed by ultrafast Kerr rotation"
}

\author{Kazuhiro Kikuchi}
\email{s2320272@u.tsukuba.ac.jp}
\affiliation{Department of Applied Physics, Graduate school of Pure and Applied Sciences, University of Tsukuba, 1-1-1 Tennodai, Tsukuba 305-8573, Japan}
\author{Yu Mizukoshi}
\affiliation{Department of Applied Physics, Graduate school of Pure and Applied Sciences, University of Tsukuba, 1-1-1 Tennodai, Tsukuba 305-8573, Japan}
\author{Takumi Fukuda}
\affiliation{Department of Applied Physics, Graduate school of Pure and Applied Sciences, University of Tsukuba, 1-1-1 Tennodai, Tsukuba 305-8573, Japan}
\affiliation{Femtosecond Spectroscopy Unit, Okinawa Institute of Science and Technology Graduate University, 1919-1 Tancha, Onna, Okinawa, Japan}
\author{Paul Fons}
\affiliation{Department of Electronics and Electrical Engineering, School of Integrated Design Engineering, Keio University, 4-1-1 Hiyoshi, Kohoku district, Yokohama city 223-8521, Japan}
\author{Muneaki Hase}
\email{Authors to whom correspondence should be addressed: mhase@bk.tsukuba.ac.jp}
\affiliation{Department of Applied Physics, Graduate school of Pure and Applied Sciences, University of Tsukuba, 1-1-1 Tennodai, Tsukuba 305-8573, Japan}

\date{\today}

\begin{abstract}
\end{abstract}

\maketitle

In order to clarify the crystal axis of the sample used, polarization dependent coherent phonon spectroscopy was carried out by transient reflectivity measurements. The experimental set up is shown in Figure S1(a). The light source was a Ti:Sapphire oscillator (average power 330 mW, central wavelength 830 nm, pulse duration $\sim$ 20 fs, repetition rate 80 MHz). 
The measured coherent phonon signal is shown in the inset of Figure S1(b) together with the Fourier transform (FT) spectrum in Figure S1(b). In Figure S1(b), the $\mathrm{A_{1g}^1}$ (2.16 THz) and $\mathrm{A_{1g}^2}$ modes (5.30 THz) were observed\cite{richter1977raman,BiSeRamanNanolett2011}. The pump polarization angle($\theta$) dependence of FT intensity is shown in Figure S1(c). The experimental results in Figure S1(c) was well fitted by the following equation\cite{norimatsu2015allphonon}.
\begin{equation}
a^2 + 2\phi^2 ab \sin^2 (\theta+\psi).\tag{S1}
\end{equation}
Here, $a$ and $b$ are the components of the Raman tensor, $\theta$ is the angle of pump polarization, $\phi$ is the incident angle of the pump light, and the $\psi$ is the phase of sinusoidal term which corresponds to the angle of the sample. It is assumed that the pump light incidents roughly normal to the sample surface ($\phi<<1$). From the fitted parameter $\psi$, we roughly determined the angle of the mirror plane of our sample.

\begin{figure}[p]
    \centering
    \includegraphics[width = 17cm]{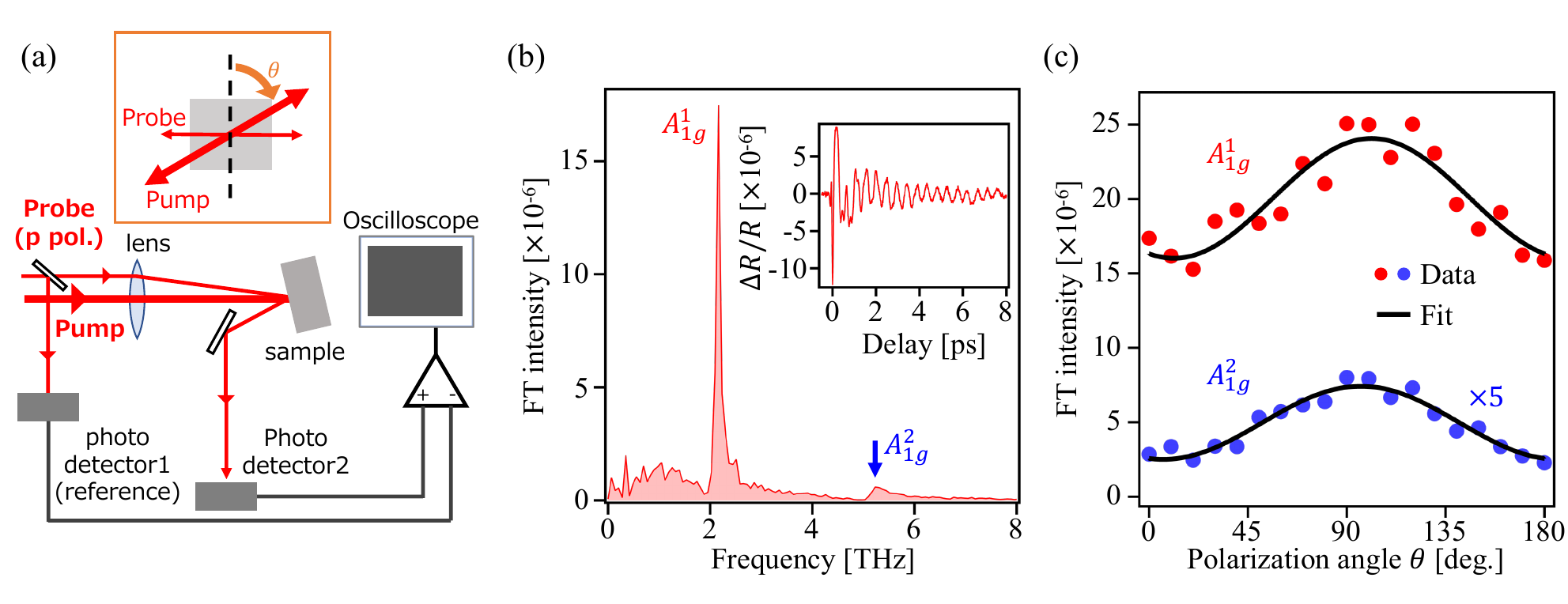}
    \caption{
    (a) Schematic of the experimental setup for transient reflectivity measurement. The inset shows the configuration of the pump and probe polarizations. (b) The obtained coherent phonon spectrum. The peak at 2.16 and 5.30 THz correspond to the $A_{1g}^{1}$ and $A_{1g}^2$ phonon modes, respectively. The inset shows a time trace of measured transient reflectivity. (c) A polarization angle $\theta$ dependence of the FT intensity of coherent phonon modes.
    }
    \label{figure1}
\end{figure}

\bibliography{supplement}